\documentclass[12pt]{article}

\usepackage[latin1]{inputenc}
\usepackage{epsfig}
\usepackage{amsbsy}
\usepackage{amsmath}
\usepackage{amssymb}

\usepackage{graphicx}
\usepackage{color}
\usepackage{enumitem}

\graphicspath{{figs/}}

\newcommand{\HD}{\textcolor{black}}

\relax
\begin{document}


%

\title{\textbf{Assessing the similarity of dose response and target doses in two non-overlapping subgroups}}

\author{Frank Bretz$^1$, Kathrin M\"ollenhoff$^{2}$,  Holger Dette$^{2}$,\\ Wei Liu$^3$, Matthias Trampisch$^{4}$}

 \maketitle

 $1$ Novartis Pharma AG, CH-4002 Basel, Switzerland \\
$2$ Department of Mathematics, Ruhr-Universit\"at Bochum, Germany \\
$3$ S3RI and School of Mathematics, University of Southampton, SO17 1TB, UK\\
$4$ Boehringer Ingelheim Pharma GmbH \& Co. KG, Biostatistics + Data Sciences / BDS, Germany
%


\begin{abstract}

We consider two problems of increasing importance in clinical dose finding studies.
First, we assess the similarity of two non-linear regression models for two non-overlapping subgroups of patients over a restricted covariate space.
To this end, we derive a confidence interval for the maximum difference between the two given models.
If this confidence interval excludes the equivalence margins, similarity of dose response can be claimed.
Second, we address the problem of demonstrating the similarity of two target doses for two non-overlapping subgroups, using again a confidence interval based approach.
We illustrate the proposed methods with a real case study and investigate their operating characteristics (coverage probabilities, Type I error rates, power) via simulation.

\end{abstract}

\noindent Keywords and Phrases: dose finding, equivalence testing, target dose estimation, subgroup analysis

\maketitle\newpage

\section{Introduction}
\label{sec:intro}


Establishing dose response and selecting optimal dosing regimens is a fundamental step in
the investigation of any new compound, be it a medicinal drug, an herbicide or fertilizer, a molecular
entity, an environmental toxin, or an industrial chemical \cite{Ruberg95a}.
This has been recognized for many years, especially
in the drug development area, where patients are exposed to a medicinal drug once it has been released
on the market.
An indication of the importance of properly conducted dose response studies is the early publication
of the tripartite ICH E4 guideline, which gives recommendations on the design and conduct of studies
to assess the relationship between doses, blood levels and clinical response throughout the clinical development of a new drug \cite{ICHE4}.

\begin{figure}[h!]
    \centering
\begin{tabular}{ll}
    $(a)$ & $(b)$ \\
    \includegraphics[width=0.5\textwidth]{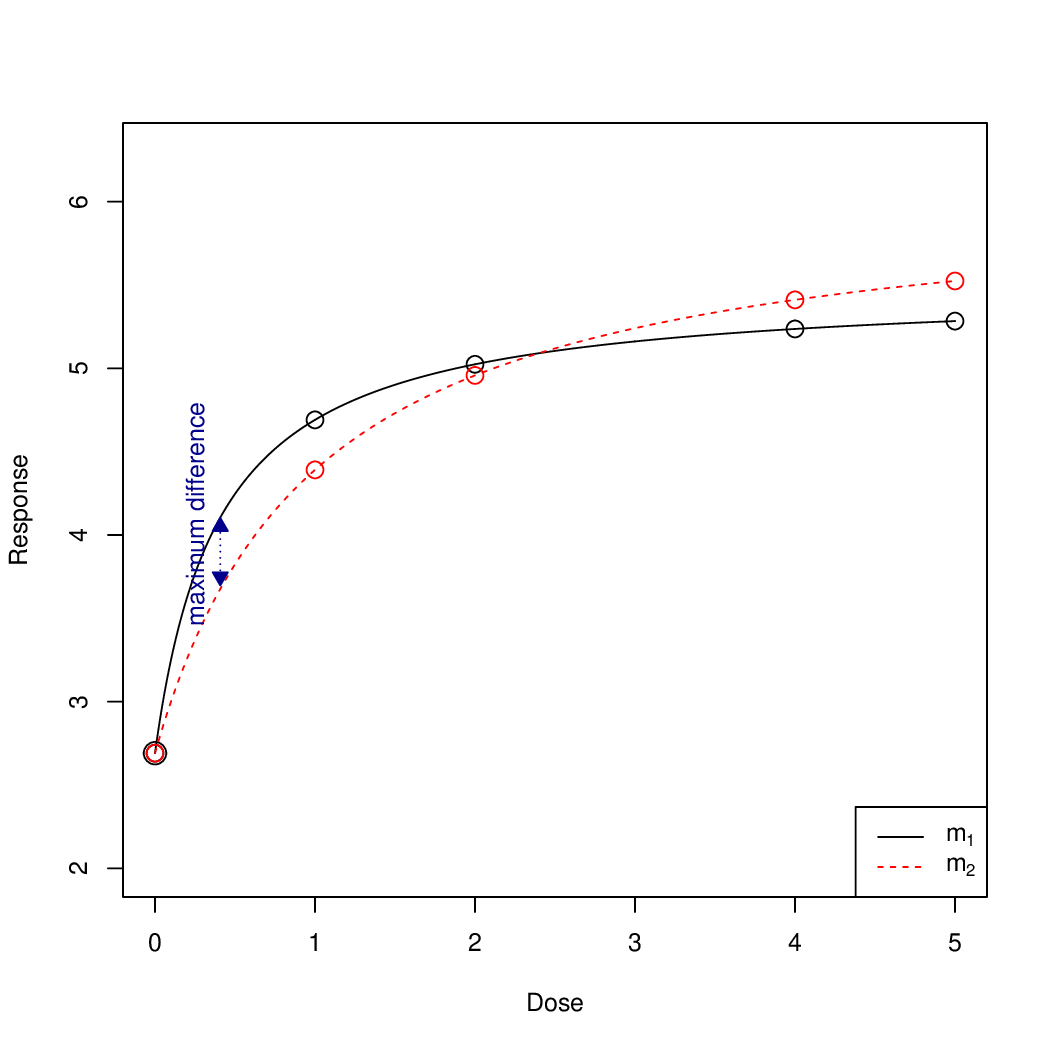} &
    \includegraphics[width=0.5\textwidth]{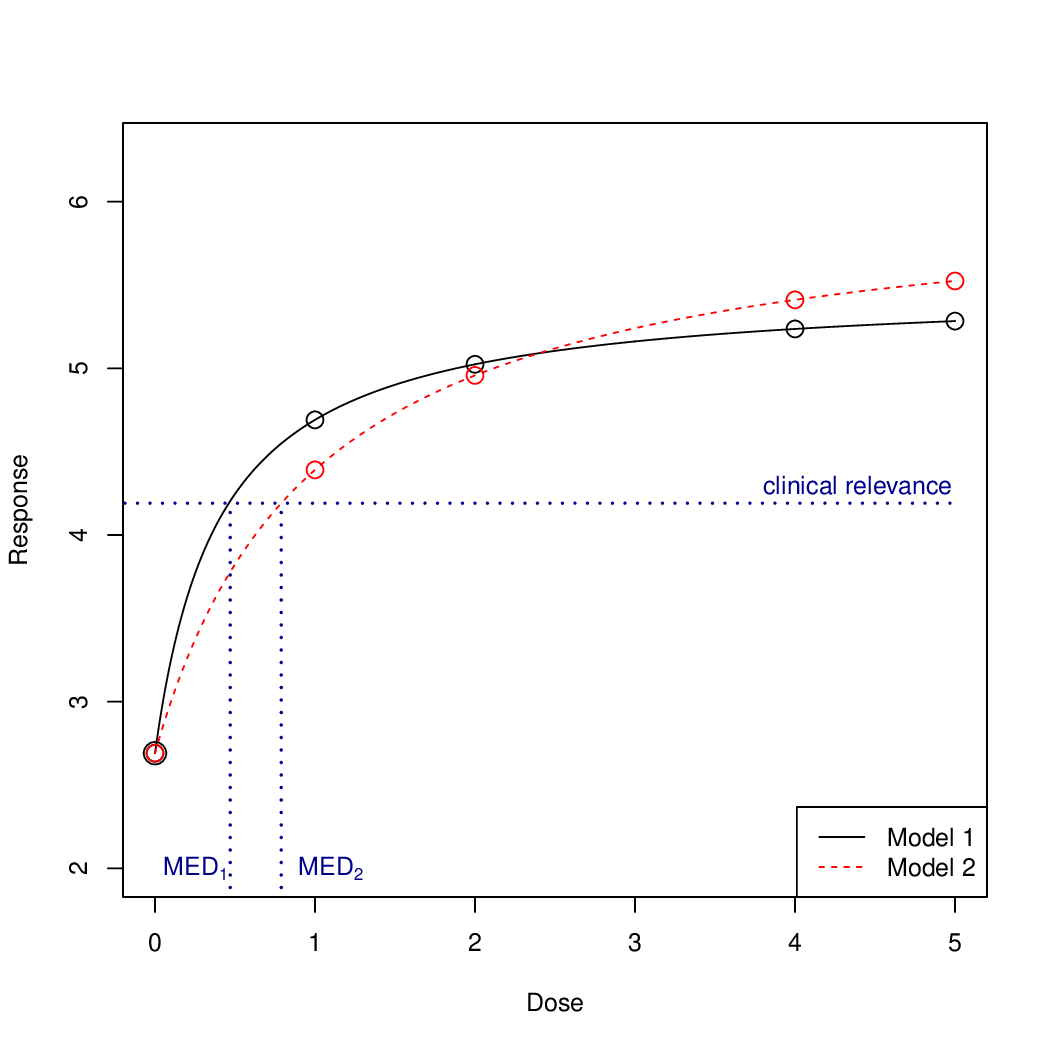}
\end{tabular}
\caption{Assessing similarity for $(a)$ two dose response curves and
$(b)$ two same target doses.} \label{fig1}
\end{figure}

Clinical trials are often analyzed beyond the primary study objectives by assessing
efficacy and safety profiles in clinically relevant subgroups, such
as different gender, age classes, grades of disease severity, etc.;
see \cite{Jhee2004,Otto2008} among many others for clinical examples.
A natural question is then whether
the dose response results are consistent across subgroups.
To illustrate the general problem, assume that we are interested in
assessing similarity for $(a)$ two dose response curves or
$(b)$ two same target doses, say for male and female patients.
For question $(a)$ we thus want to
show that the maximum difference in response between two (potentially
different) non-linear parametric regression models is smaller than a
pre-specified margin.
Figure~\ref{fig1}$a$ displays an example, where the two dose response
curves follow different Emax models. The maximum response difference
over the dose range is indicated by the arrow. For question $(b)$ we
want to show that two same target doses do not differ relevantly.
Figure~\ref{fig1}$b$
displays the minimum effective dose ($MED$) derived from the two
previous dose response models. Here, the $MED$ is defined
as the smallest dose which demonstrates a clinically relevant
benefit over placebo, as indicated by the horizontal line in Figure~\ref{fig1}$b$.
If we succeed in demonstrating either $(a)$ or
$(b)$, evidence is provided that the difference in response over
the entire dose range or the two target
doses differ at most marginally. In practice, such a result may
provide sufficient evidence that the same dose can be administered
in both subgroups (e.g. the same doses for male and female patients).

In this paper we focus on model-based approaches for Phase II dose finding trials. Compared to traditional
analysis-of-variance (ANOVA) approaches based on pairwise multiple comparisons, they have the advantage of enabling
the use of more doses in the design, without requiring a larger number of patients. In an ANOVA-type approach
only the information from the dose levels under investigation is used to declare a dose response signal. Consequently, the required
sample size depends strongly on the number of dose levels under investigation when a fixed precision is required at each dose level.
Modeling techniques allow one to interpolate information across dose and the total sample size will depend less strongly
on the number of dose levels under investigation. The possibility of using more dose levels will typically result in information-richer
trial designs and a better basis for decision making at the end of Phase II.
This has been confirmed by several simulation studies in the literature, such as the
White Paper of the PhRMA working group on  ``Adaptive Dose-Ranging Studies'' \cite{Bornkamp2007}.
The main objective of this group was to evaluate different novel and existing model-based dose ranging methods
in a comprehensive simulation study, as compared to an ANOVA approach. In summary, one can conclude
from the PhRMA simulations that model-based methods outperformed the benchmark ANOVA approach in many cases.
In the meantime, the use of model-based approaches in Phase II dose finding trials
has been supported by several major regulatory agencies \cite{EMA2015}.

As insinuated by Figure~\ref{fig1}, Phase II dose finding trials have multiple, concurrent objectives \cite{Ruberg95a,Bretz08}.
A common objective is to give a complete functional description of the dose response relationship.
An alternative objective is to estimate a target dose for the subsequent confirmatory Phase III trials.
However, demonstrating similarity of target doses or dose
response curves in each of several subgroups
has not been addressed in much detail so far in the literature.
One exception is \cite{Dette2016}, who proposed a non-standard bootstrap approach for
question $(a)$ which addresses the specific form of the interval hypotheses.
In particular, data has to be generated under the null hypothesis using constrained least squares estimates.
In this paper we consider different methods to address both
questions $(a)$ and $(b)$.
Extending the work from \cite{Jin16} and using the results from \cite{Liu09},
we address problem $(a)$  in Section~\ref{sec:dr} by
deriving a confidence interval for the maximum difference between the two given
non-linear regression models over the entire covariate space of interest.
If this confidence
interval excludes the equivalence margins,
similarity of dose response can be claimed. In Section~\ref{sec:td}, we
consider asymptotic methods
to derive confidence intervals for the
difference between two same target doses to address problem $(b)$.
Again, if such a confidence interval excludes a pre-specified relevance margin,
similarity in dose can be claimed.
In Section~\ref{sec:conc} we provide some concluding remarks.
Technical details are left for the Appendix.

\section{Assessing similarity of two dose response curves }
\label{sec:dr}

We consider the non-linear regression models
\begin{equation}\label{reg_model}
 Y_{\ell ,i,j} = m_\ell(\vartheta_\ell,d_{\ell ,i})+\epsilon_{\ell ,i,j}~,~j=1,\ldots , n_{\ell ,i},~i=1, \ldots, k_\ell,\ \ell=1,2 ,\ d_{\ell ,i}\in\cal D,
\end{equation}
where $Y_{\ell,i,j}$ denotes the $j$th observed response at the $i$th dose level $d_{\ell,i}$ under the $\ell$th dose response model $m_\ell$.
The error terms $\epsilon_{\ell,i,j}$ are assumed to be independent and identically distributed with expectation $0$ and variance $\sigma_\ell^2$.
Further, $n_\ell =\sum_{i=1}^{k_\ell}n_{\ell ,i}$ denotes the sample size in group $\ell$ where we assume $n_{\ell,i}$ observations in the $i$th dose level ($i=1,\ldots k_\ell,\ \ell=1,2)$.
We further assume that for both regression models the different dose levels
are attained on the same (restricted) covariate region $\cal D$.
For the purpose of this paper, we assume $\cal D$ to be the dose range
under investigation, although the results in this section
can be generalized to include other covariates.
The functions $m_1$ and $m_2$ in (\ref{reg_model}) denote the (non-linear) regression models with fixed
but unknown $p_1$-  and $p_2$-dimensional parameter vectors $\vartheta_{1}$ and $\vartheta_{2}$, respectively.
Note that both the regression models $m_1$
and $m_2$ and the parameters $\vartheta_1$ and $\vartheta_2$ may be different.
In particular, the design matrices for the two regression
models may be unequal. This implies that we do not assume
the same doses to be investigated for $\ell=1,2$ and that the sample sizes $n_\ell$ can be
unequal. We refer to \cite{Pinheiroetal06} for an overview of several
linear and non-linear regression models commonly employed in
clinical studies.

\subsection{Methodology}
\label{ssec:meth1}

Using results from \cite{Liu09}, we
derive in the following a confidence
interval for the maximum absolute difference between the two given non-linear
regression models $m_1$ and $m_2$ over the entire covariate space $\cal D$.
We use this confidence interval in order to derive a test demonstrating similarity of the two dose response curves.

Let $U\left(Y_{1},Y_{2},d\right)$ denote a $1-\alpha$ pointwise
upper confidence bound on the difference curve $m_{2}(\mathbf{\vartheta}_{2},d)-m_{1}(\mathbf{\vartheta}_{1},d)$, i.e.
$P\left\{ m_{2}(\mathbf{\vartheta}_{2},d)-m_{1}(\mathbf{\vartheta}_{1},d)
\le U\left(Y_{1},Y_{2},d\right)\right\} \ge1-\alpha$ for all $d\in {\cal D}$, where
$\alpha$ denotes the pre-specified significance level and $ Y_{\ell}$ the vector of observations from group $\ell=1,2$.
Similarly, let $L\left(Y_{1},Y_{2},d\right)$ denote a $1-\alpha$ pointwise
lower confidence bound on $m_{2}(\mathbf{\vartheta}_{2},d)-m_{1}(\mathbf{\vartheta}_{1},d)$.
Using these pointwise confidence bounds we can deduce a confidence interval for the maximum absolute difference between the two models $\max_{d\in{\cal D}} |m_{2}(\mathbf{\vartheta}_{2},d)-m_{1}(\mathbf{\vartheta}_{1},d)|$ over the region $\cal D$, that is
\begin{equation}\label{conf_band}
P\left\{ \max_{d\in{\cal D}} |m_{2}(\mathbf{\vartheta}_{2},d)-m_{1}(\mathbf{\vartheta}_{1},d)|
\leq \max \big\{ \max_{d\in{\cal D}} U\left(Y_{1},Y_{2},d\right),-\min_{d\in{\cal D}} L\left(Y_{1},Y_{2},d\right)\big\} \right\} \geq 1-\alpha.
\end{equation}
The proof is given in Appendix~A.
\HD{
For moderate sample sizes the pointwise confidence bounds $U\left(Y_{1},Y_{2},d\right)$
and $L\left(Y_{1},Y_{2},d\right)$ can be derived from the delta method  \cite{Serfling1980}.
Let $u_{1-\alpha}$ denote the $1-\alpha$ quantile of the standard normal distribution. Then,
\begin{equation*}\label{eq:pointwise}
U\left(Y_{1},Y_{2},d\right)
= m_{2}(\hat{\vartheta}_{2},d)-m_{1}(\hat{\vartheta}_{1},d) + u_{1-\alpha} \hat\rho(d)
\end{equation*}
and
\begin{equation*}
L\left(Y_{1},Y_{2},d\right)
= m_{2}(\hat{\vartheta}_{2},d)-m_{1}(\hat{\vartheta}_{1},d) - u_{1-\alpha} \hat\rho(d)
\end{equation*}
are the desired $1-\alpha$ asymptotic pointwise upper and lower confidence bounds, respectively,  for
$m_{2}(\mathbf{\vartheta}_{2},d)-m_{1}(\mathbf{\vartheta}_{1},d)$. Here,
$\hat{\vartheta}_{\ell}$ denotes the least squares estimate  of $\vartheta_{\ell}$
and
\begin{equation} \label{varest}
\hat \rho^2(d) = \frac{\hat \sigma_1^2}{n_1} \big(
\tfrac{\partial} {\partial \vartheta_1}   m_1(  \hat \vartheta_1,d)  \big)^T     ~ \hat \Sigma_1^{-1}  ~  \big(
\tfrac{\partial} {\partial \vartheta_1}   m_1(  \hat \vartheta_1,d)  \big)
 + \frac{\hat \sigma_2^2}{n_2}  \big(
\tfrac{\partial} {\partial \vartheta_2}   m_2(  \hat \vartheta_2,d)  \big)^T ~ \hat \Sigma_2^{-1}  ~\big ( \tfrac{\partial} {\partial \vartheta_2}   m_2(  \hat \vartheta_2,d)  \big)
\end{equation}
is an estimate of the variance of  $m_{2}(\hat{\vartheta}_{2},d)-m_{1}(\hat{\vartheta}_{1},d) $. In \eqref{varest}
$\hat\sigma_\ell^2$, is the common variance estimate in the $\ell$th group  ($\ell=1,2$) and
 $\hat\Sigma_{\ell}= \sum_{i=1}^{k_\ell}\tfrac{n_{\ell,i}}{n_\ell}\tfrac {\partial}{\partial \vartheta_\ell}  m_\ell (x_{\ell,i,}, \hat \vartheta_\ell )
 \big(\tfrac {\partial}{\partial \vartheta_\ell}  m_\ell (x_{\ell,i,}, \hat \vartheta_\ell )\big)^T$. Note that the matrix
 $\frac{\hat \sigma_\ell^2}{n_\ell}\hat\Sigma_{\ell}^{-1} $
  is a consistent estimator of the covariance matrix of $\hat \vartheta_\ell$ ($\ell=1,2$).}

Next we are interested in demonstrating that the maximum absolute
difference in response between the two regression models in
(\ref{reg_model}) over the covariate space $\cal D$
is not larger than a
pre-specified margin $\delta>0$.
Formally, we test the null hypothesis
\begin{eqnarray}\label{H0}
H:\max_{d\in{\cal D}}\left | m_{2}(\mathbf{\vartheta}_{2},d)-m_{1}(\mathbf{\vartheta}_{1},d)\right | \ge\delta
\end{eqnarray}
against the alternative hypothesis
\begin{eqnarray}\label{H1}
K:\max_{d\in{\cal D}}\left | m_{2}(\mathbf{\vartheta}_{2},d)-m_{1}(\mathbf{\vartheta}_{1},d)\right | <\delta.
\end{eqnarray}
Consequently, using the confidence interval \eqref{conf_band}, equivalence is claimed if
\begin{equation*}
\max \big\{ \max_{d\in{\cal D}} U\left(Y_{1},Y_{2},d\right),-\min_{d\in{\cal D}} L\left(Y_{1},Y_{2},d\right)\big\} < \delta.
\end{equation*}
Thus, we reject the null hypothesis $H$ at level $\alpha$ and assume similarity of $m_1$ and $m_2$ if
\begin{equation}\label{reject_H}
-\delta < \min_{d\in\cal D} L\left(Y_{1},Y_{2},d\right)
\quad \mbox{and} \quad  \max_{d\in\cal D} U\left(Y_{1},Y_{2},d\right) < \delta.
\end{equation}

\subsection{Case study}
\label{ssec:case1}

To illustrate the methodology described in Section~\ref{ssec:meth1},
we consider a dose finding trial for a weight loss drug given
to patients suffering from overweight or obesity.
This trial aims at comparing the dose response relationship for two regimens, namely a once-daily (o.d.) and a twice-daily (b.i.d.) application of the drug.
The primary objective in this trial is not to apply a joint model that includes both regimen, but rather
treat both regimen separately and assess the similarity of dose response.
Because this study has not been completed yet, we simulate data based on the assumptions made at the
trial design stage. For confidentiality reasons, we use blinded dose levels and
all chosen dose levels denote the total daily dose. These limitations do not change the utility of the calculations below.

In this trial, the dose levels for the o.d. and b.i.d. regimens are given by $0.033,\ 0.1, 1$
and $0.067,\ 0.3,\ 1$, respectively.
Patients are thus randomized to receive either placebo or one of the six active treatments.
In total, we assume that 350 patients are allocated equally across the seven arms,
resulting in a sample size of $50$ patients per treatment group.
The primary endpoint of the study was the percentage of weight loss after a treatment duration of 20 weeks,
with smaller values corresponding to a better treatment effect.

We used the \texttt{nls} function in \texttt{R} \cite{r:2015} to compute the
non-linear least squares estimates $\hat{\vartheta}_{\ell}$ of $\vartheta_{\ell}$
and the standard errors necessary for calculating
$U\left(Y_{1},Y_{2},d\right)$
and $L\left(Y_{1},Y_{2},d\right)$
from Section 2.1. The \texttt{R} code for this example and
all other calculations in this paper is available from
the authors upon request.

\begin{figure}[h!]
    \begin{center}
\begin{tabular}{ll}
    $(a)$ & $(b)$ \\
    \includegraphics[width=0.5\textwidth]{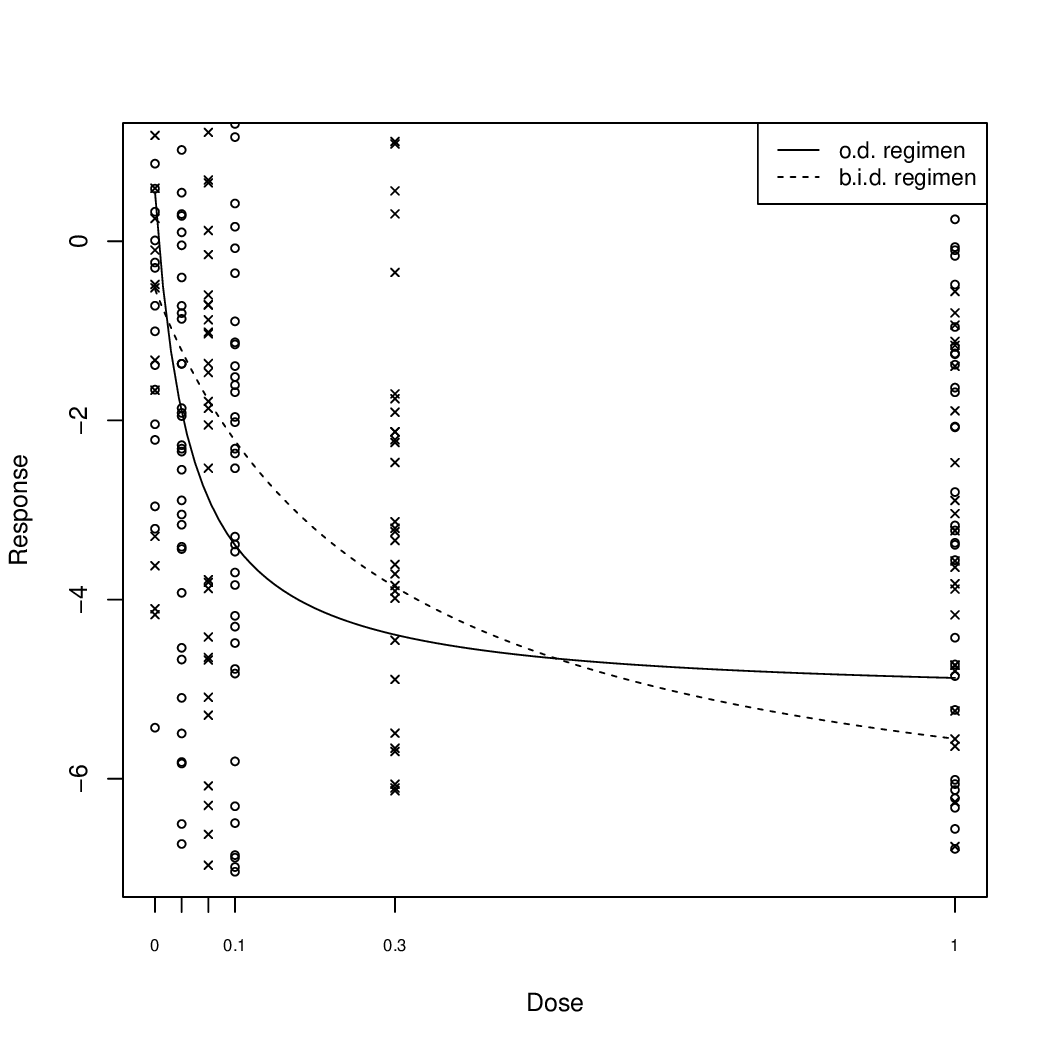} &
    \includegraphics[width=0.5\textwidth]{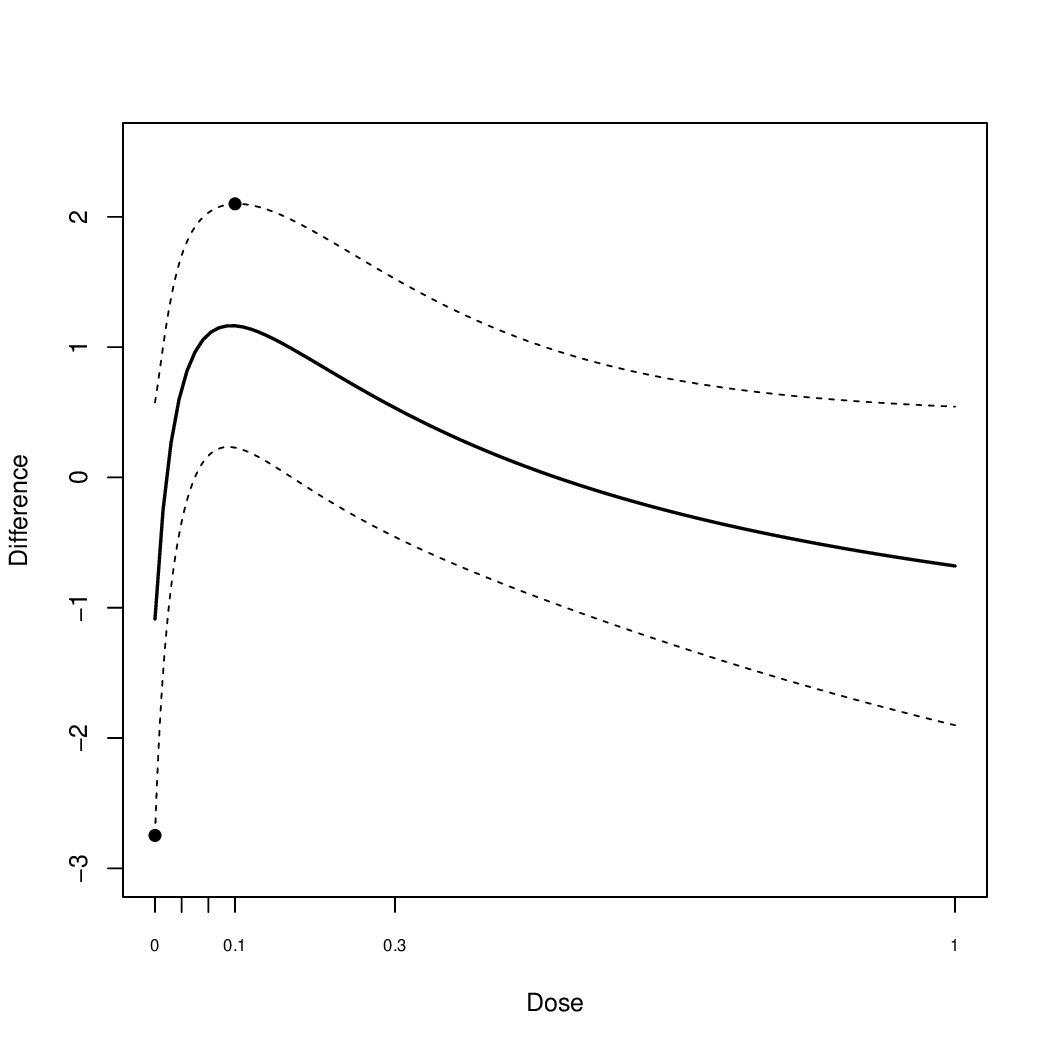}
\end{tabular}
    \end{center}
\caption{Plots for the weight loss case study.
$(a)$ The fitted Emax model $m_1$ ($m_2$) for the o.d. (b.i.d.) regimen is given by the solid (dashed) line with observations marked by ``x'' (``o'').
$(b)$ Mean difference curve with associated pointwise 90\% confidence bounds. Bold dots denote the maximum upper and minimum lower confidence bound over ${\cal D} = [0,1]$.}
\label{fig2}
\end{figure}

For this example, we fitted two Emax models:
$m_1(\vartheta_1, d) = \vartheta_{1,1} + \vartheta_{1,2}\frac{d}{\vartheta_{1,3} + d}$ for the o.d. regimen
and
$m_2(\vartheta_2, d) = \vartheta_{2,1} + \vartheta_{2,2}\frac{d}{\vartheta_{2,3} + d}$ for the b.i.d. regimen,
where $\vartheta_1 = (\vartheta_{1,1}, \vartheta_{1,2}, \vartheta_{1,3})$ and
$\vartheta_2 = (\vartheta_{2,1}, \vartheta_{2,2}, \vartheta_{2,3})$.
For the data set at hand, $\hat{\vartheta}_{1} =(0.55,-5.66, 6.55)$
and $\hat{\vartheta}_{2} = (-0.54,-6.42, 41.99)$.
Figure~\ref{fig2}$a$ displays the fitted dose response
models $m_1(\hat{\vartheta}_{1}, d)$ and $m_2(\hat{\vartheta}_{2}, d)$, $d \in [0, 1]$,
together with the individual observations, where the vertical axis
is truncated to $[-7, 1]$ for better readability.
Figure~\ref{fig2}$b$ displays
the difference $m_2(\hat{\vartheta}_{2}, d) - m_1(\hat{\vartheta}_{1}, d)$
together with the associated 90\% pointwise confidence intervals for each dose
$d \in [0, 1]$.
The maximum upper confidence bound for $\alpha=0.1$ is
$\max_{d\in\cal D}U\left(Y_{1},Y_{2},d\right)=2.099$
at dose $d=0.1$ and the minimum lower confidence bound is
$\min_{d\in\cal D} L\left(Y_{1},Y_{2},d\right)=-2.748$ at the minimum dose $d=0$.
That is, the maximum difference in response between the two regimens over the dose range
$\mathcal{D}=\left[0,1\right]$ lies between $-2.748$ and $2.099$. Therefore, similarity  of the dose response curves
can be claimed at level $\alpha=0.1$
as long as $\delta$ is larger than $2.748$, according to (\ref{reject_H}).


\subsection{Simulations}
\label{ssec:sim1}

We conducted a simulation study to investigate the operating
characteristics of the method described in Section~\ref{ssec:meth1}.
We investigated coverage probabilities of the confidence intervals as well as Type I error rates and
power of the test~(\ref{reject_H}) for different scenarios.
To simplify the simulations, we assumed balanced designs and that dose is the only
covariate. For all simulations below, we generated data as follows:
\begin{enumerate}[leftmargin=1.3cm]
\item[\textbf{Step 1:}] Specify the models $m_{1}, m_{2}$, their parameters $\vartheta_{1}, \vartheta_{2}$, a common variance
$\sigma^2$ and the actual dose levels $d_{\ell,i}$.
\item[\textbf{Step 2:}] Generate $n_{\ell,i}$ values $m_{\ell}(\vartheta_{\ell},d_{\ell,i})$ at each dose $d_{\ell,i}$.
\item[\textbf{Step 3:}] Generate normally distributed residual errors $\epsilon_{\ell,i,j}\sim N(0,\sigma^{2})$ and use the final response data
\begin{equation}
Y_{\ell,i,j} = m_{\ell}(\mathbf{\vartheta}_{\ell},d_{\ell,i})+\epsilon_{\ell,i,j},\qquad j=1, \ldots, n_{\ell,i},\ i=1, \ldots k_\ell,\ \ell=1,2.
\label{algorithmus}
\end{equation}
\end{enumerate}
This procedure is repeated using $10,000$ simulation runs.
Because of the large number of scenarios, only a subset of the possible results is included below to illustrate the key findings. The complete simulations results are available in \cite{Bretz2016}.

\subsubsection{Coverage probabilities}
\label{sssec:cp}

In the following we report the coverage probabilities of the confidence intervals for the maximum absolute difference derived in \eqref{conf_band} under two different scenarios.

\begin{figure}[h!]
    \begin{center}
\begin{tabular}{cc}
    $(a)$ Scenario 1 & $(b)$ Scenario 2\\
    \includegraphics[width=0.5\textwidth]{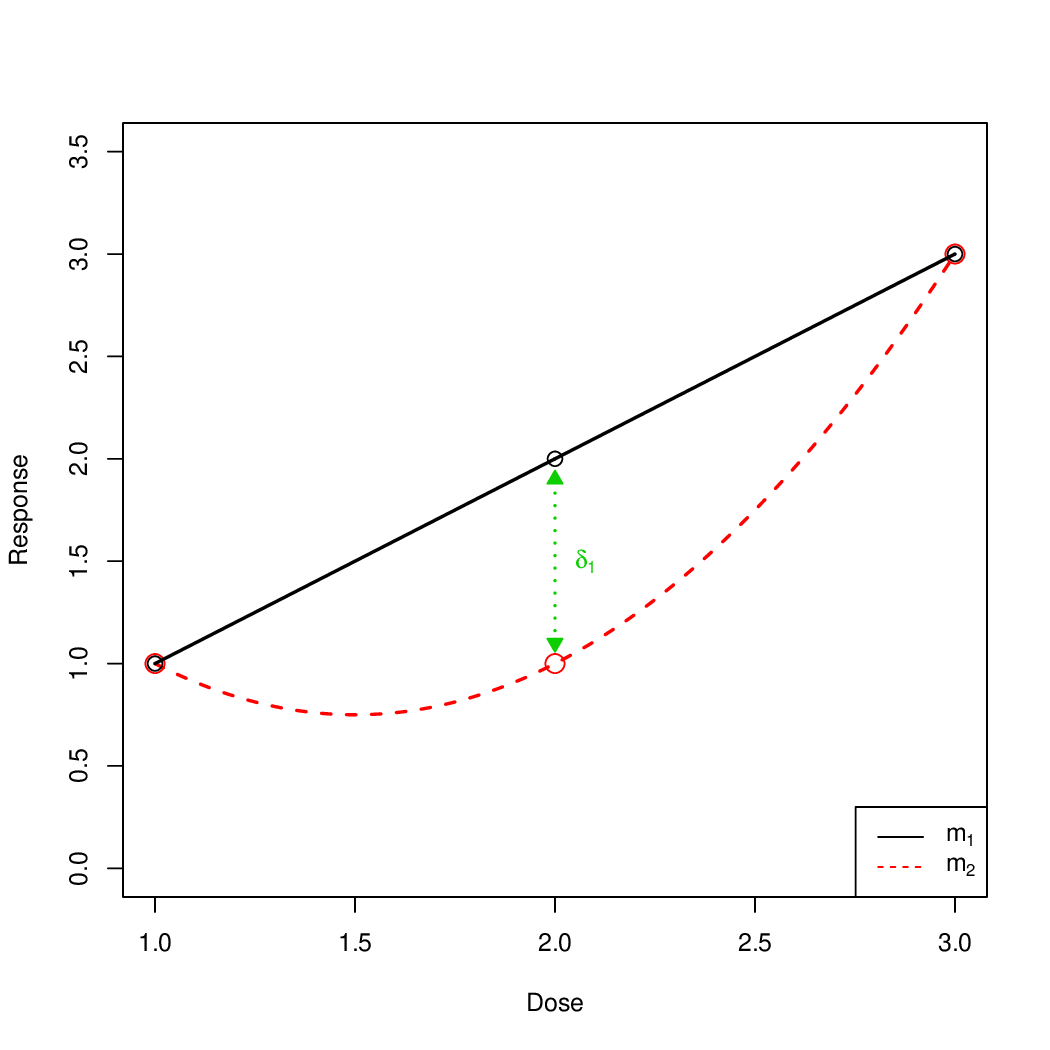} &
    \includegraphics[width=0.5\textwidth]{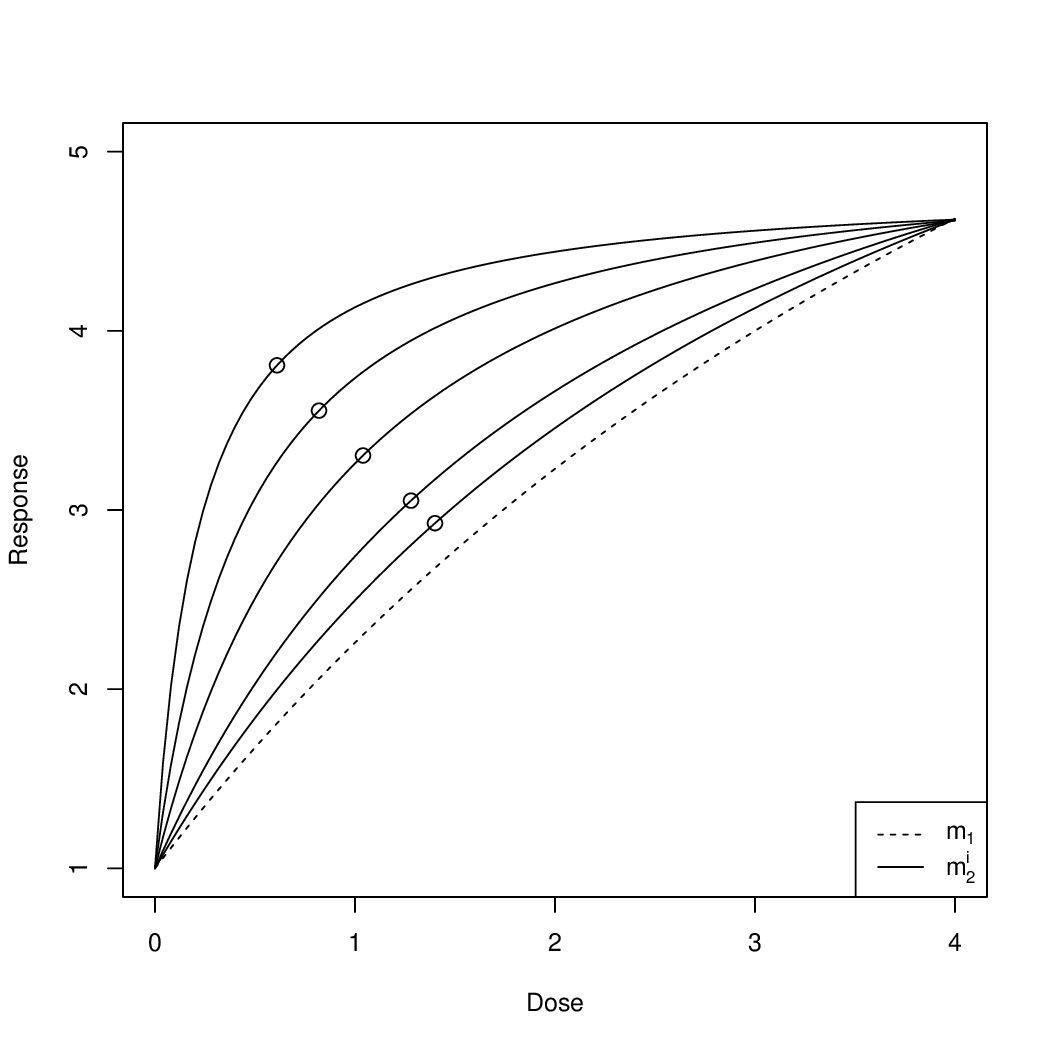}
\end{tabular}
    \end{center}
\caption{Graphical illustration of the two scenarios used for the simulations.
Open dots in the left panel indicate the actual dose levels.
In the right panel they indicate the doses where the maximum distance to the reference curve $m_1$ (dashed line) is observed.}
\label{fig3}
\end{figure}

\paragraph{Scenario 1}
We start with the comparison of a linear and a quadratic model.
More specifically, we chose the linear model $m_1(d)=d$ and
the quadratic model $m_2(d)=3\delta_1+(1-4\delta_1)d+\delta_1 d^{2}$, $d \in [1,3]$;
see Figure~\ref{fig3}$a$ for $\delta_1=1$.
We assumed identical dose levels $d_{\ell,i} = i$, $i=1, 2, 3$
for both regression models $\ell = 1, 2$.
Consequently, the two curves
coincide at the two boundary doses $d = 1, 3$, and the maximum difference $\delta_1$ occurs
at dose $d=2$.
For each configuration of $\sigma^2 = 1, 2, 3$ and $\delta_1=1, 2, 3$
we used \eqref{algorithmus} to simulate
$n_{\ell,i}=10 (50)$ observations at each dose level $d_{\ell,i}$, resulting in $n_{\ell} = 30 (150)$, $\ell = 1, 2$.

The left side of Table~\ref{tab1} displays the coverage probabilities for $\alpha = 0.05, 0.1$.
We observe that the nominal level of $1-\alpha$ is reached in all cases under consideration, which confirms \eqref{conf_band}.
The confidence intervals
are more accurate for larger sample sizes and smaller variances, because we used the asymptotic quantiles from the normal distribution. If, instead,
we select the quantiles from the $t$ distribution, the simulated coverage probabilities are closer to the nominal $1-\alpha$ level (results not shown here).
Note that the confidence bounds perform better for larger values of $\delta_1$.
\HD{This effect
 can be explained by a careful look at the proof given in Appendix~A and the particular example under consideration.
 First note that the maximum absolute difference $\delta_1$ between the two curves is attained at a single point, say $d_0$; see Figure~\ref{fig3}$a$. If this
 difference is large then either $ \max_{d\in{\cal D}} U\left(Y_{1},Y_{2},d\right) = U\left(Y_{1},Y_{2},d_0\right) $
 or $-\min_{d\in{\cal D}} L\left(Y_{1},Y_{2},d\right)  = L\left(Y_{1},Y_{2},d_0\right)$  with high probability    and consequently there is
 equality either in \eqref{a1} or \eqref{a2} in Appendix A. The same effect appears for increasing sample sizes and
 smaller values of $\delta_1$ as in this case the parameter estimates and approximation of the coverage probability of the confidence interval
are more precise.}

\begin{table}[h!]
\begin{center}
\makebox[\linewidth]{ 
\tiny{\begin{tabular}{cc|ccc|ccc|ccc|ccc} \hline\hline
& & \multicolumn{6}{c}{Coverage probabilities} \vline & \multicolumn{6}{c}{Type I error rates}\\
\cline{3-14}
& & \multicolumn{3}{c}{$\alpha=0.05$} &  \multicolumn{3}{c}{$\alpha=0.1$} \vline & \multicolumn{3}{c}{$\alpha=0.05$} & \multicolumn{3}{c}{$\alpha=0.1$}\\ \cline{3-14}
$\delta_1$ & $\sigma^{2}$ & $n_{\ell}=30$ & $n_{\ell}=90$ & $n_{\ell}=150$ &  $n_{\ell}=30$ & $n_{\ell}=90$ & $n_{\ell}=150$ & $n_{\ell}=30$ & $n_{\ell}=90$ & $n_{\ell}=150$ &  $n_{\ell}=30$ & $n_{\ell}=90$ & $n_{\ell}=150$ \\ \hline
   1 &   1 & 0.987 & 0.950 &  0.950 &  0.953 & 0.915 & 0.906 & 0.012 &0.050 & 0.050&0.046 & 0.085 &0.095 \\
   1 &   2 & 0.999 & 0.973 & 0.956 &  0.991 & 0.929 &0.906 & 0.001   &0.027 & 0.042&0.009 & 0.071 &0.088 \\
   1 &   3 & 1.000 & 0.992 & 0.971 &  0.999 & 0.965 &0.923 & 0.000   &0.008 & 0.031&0.001 & 0.035 &0.077 \\
   2 &   1 & 0.949 & 0.942& 0.952 &  0.901 & 0.909 &0.907 & 0.047    & 0.058& 0.049 &0.096 & 0.091&0.105 \\
   2 &   2 & 0.960 & 0.959 & 0.951 & 0.913 & 0.911 &0.901 & 0.039    & 0.051& 0.048 &0.079 & 0.089&0.095 \\
   2 &   3 & 0.977 & 0.946 &0.950  & 0.936 & 0.902 &0.902 & 0.025    & 0.054& 0.047 &0.065 & 0.099&0.097 \\
   3 &   1 & 0.951 &  0.945 &0.954  & 0.906 & 0.904 &0.908 & 0.053   &0.055 & 0.048 &0.102 & 0.096&0.100 \\
   3 &   2 & 0.952 & 0.941 &0.954  & 0.905 & 0.895 &0.907 & 0.048    &0.059 & 0.047&0.094 & 0.105 &0.099 \\
   3 &   3 & 0.949 & 0.946 & 0.952  & 0.900 & 0.902 &0.903 & 0.052   & 0.054& 0.049 &0.098 & 0.098&0.099 \\
\hline\hline
\end{tabular}}}
\end{center}
\caption{Simulated coverage probabilities and Type I error rates
for different configurations of $\delta_1$, $\sigma^{2}$, $\alpha$, and $n_{\ell}$ under Scenario 1.}
\label{tab1}
\end{table}

\paragraph{Scenario 2}
We now consider the comparison of two different Emax models,
where the maximum distances with respect to the same reference model are $0.25,\ 0.5,\ 1,\ 1.5$ and $2$.
More specifically, we compared the reference Emax model $m_1(d)=1+\frac{9.70d}{6.70+d}$ with
\footnotesize \begin{equation} m_2^1(d)=1+\frac{6.88d}{3.60+d},\ \ m_2^2(d)=1+\frac{5.66d}{2.25+d},\ m_2^3(d)=1+\frac{4.52d}{1+d},
\ m^4_2(d)=1+\frac{4.05d}{0.48+d},\ m_2^5(d)=1+\frac{3.82d}{0.22+d},
\label{scenario2}
\end{equation}
\normalsize
where the dose range is given by $\mathcal{D}=[0,4]$. Note that the placebo response at $d=0$ is $1$ and the response at the highest dose $d=4$ is $4.62$ for all five models; see Figure \ref{fig3}$b$.
The difference curve is given by $m_{2}^h(\mathbf{\vartheta}^h_{2},d)-m_{1}(\mathbf{\vartheta}_{1},d)$ for $h=1,2,3,4,5$.
Note that the dose which produces the maximum difference is different for each $h$.
More precisely, these doses are given by $1.4,\ 1.28,\ 1.04,\ 0.82$ and $0.61$ for $h=1,\ldots,5$; see again Figure \ref{fig3}$b$.
The maximum absolute distance attained at each of these doses is
denoted by $\delta_\infty=\max_{d\in{\cal D}}\left | m_{2}^h(\mathbf{\vartheta}_{2}^h,d)-m_{1}(\mathbf{\vartheta}_{1},d)\right |$.
We assumed identical dose levels $d_{\ell,i} = i-1$, $i=1,2,3,4,5$
for both regression models $\ell = 1, 2$.
For each configuration of $\sigma^2 = 1, 2, 3$ and $\delta_\infty = 0.25, 0.5, 1, 1.5, 2$,
we used \eqref{algorithmus} to simulate
$n_{\ell,i}=30$ observations at each dose level $d_{\ell,i}$, resulting in $n_{\ell} = 150$, $\ell = 1, 2$.

The left side of Table~\ref{tab2} displays the coverage probabilities for $\alpha = 0.05, 0.1$.
As already seen under Scenario 1, the confidence intervals are more accurate for smaller variances (and larger sample sizes, results not shown here)
and for increasing values of $\delta_\infty$. As before, asymptotically the coverage probability is at least $1-\alpha$ under all configurations investigated here.

\begin{table}[h!]
\begin{center}
\makebox[\linewidth]{ 
\tiny{\begin{tabular}{ccc|ccc|ccc|ccc|ccc} \hline\hline
		& && \multicolumn{6}{c}{Coverage probabilities} \vline & \multicolumn{6}{c}{Type I error rates}  \\
		\cline{4-15}
		& & & \multicolumn{3}{c}{$\alpha=0.05$} &  \multicolumn{3}{c}{$\alpha=0.1$} \vline & \multicolumn{3}{c}{$\alpha=0.05$} & \multicolumn{3}{c}{$\alpha=0.1$}\\ \cline{4-15}
		$(m_1,m_2)$& $\delta_\infty$ & $\sigma^{2}$ & $n_{\ell}=30$ & $n_{\ell}=90$ & $n_{\ell}=150$ &  $n_{\ell}=30$ & $n_{\ell}=90$ & $n_{\ell}=150$ & $n_{\ell}=30$ & $n_{\ell}=90$ & $n_{\ell}=150$ &  $n_{\ell}=30$ & $n_{\ell}=90$ & $n_{\ell}=150$ \\ \hline
 $(m_1,m_2^1)$ & 0.25 &   1 & 1.000& 1.000&1.000 & 1.000& 1.000&1.000 & 0.000& 0.000& 0.000 &0.000 & 0.000 &0.000   \\
 &&                                   2 & 1.000& 1.000&1.000 & 1.000& 1.000& 1.000 & 0.000& 0.000& 0.000 &0.000 & 0.000 & 0.000  \\
 &&                                   3 & 1.000& 1.000&1.000 & 1.000& 1.000&1.000  & 0.000&0.000& 0.000 &0.000 & 0.000 & 0.000  \\\hline
$(m_1,m_2^2)$ &  0.5 &    1 & 1.000 & 1.000&0.994 & 1.000& 0.990&0.960 & 0.000&0.000&0.006 & 0.000& 0.001 &0.040 \\
&&                                    2 &1.000 & 1.000&1.000 & 1.000& 1.000&0.993  &0.000 &0.000&0.000 & 0.000& 0.000 &0.007  \\
&&                                    3 & 1.000& 1.000&1.000 & 1.000& 1.000&1.000  & 0.000&0.000&0.000 & 0.000&0.000  &0.000 \\\hline
$(m_1,m_2^3)$ &  1 &      1  &0.995 &  0.957 &0.954 & 0.980& 0.902 &0.893  & 0.005 &0.042&0.036 & 0.002 & 0.097 &0.107 \\
&&                                   2 & 1.000& 0.981 &0.963 & 1.000& 0.936 &0.903  & 0.000&0.019 &0.047 & 0.000 & 0.064 &0.097  \\
&&                                   3 & 1.000& 0.996 &0.983 & 1.000& 0.968 &0.942  & 0.000& 0.004 &0.015 & 0.000 & 0.031 &0.058  \\\hline
$(m_1,m_2^4)$  & 1.5 &    1 & 0.971& 0.939 &0.952 & 0.921& 0.868 &0.899  & 0.029 & 0.061 &0.048 & 0.078& 0.131 &0.101  \\
&&                                   2 & 0.996& 0.961 &0.962 & 0.966& 0.910 &0.913  & 0.004 & 0.038 &0.038 & 0.033& 0.090 &0.087  \\
&&                                   3 & 1.000& 0.965 &0.949 & 0.987& 0.907 &0.897  & 0.000& 0.035 &0.051 & 0.012  & 0.092 &0.103  \\\hline
$(m_1,m_2^5)$  & 2 &      1 & 0.940 & 0.929 &0.945 & 0.897& 0.867 &0.902  & 0.060 & 0.071 &0.055 & 0.102& 0.132&0.098  \\
&&                                   2 & 0.958 & 0.940 &0.942 &0.903 & 0.878 &0.889  & 0.041 & 0.060 &0.068 & 0.096& 0.122&0.118  \\
&&                                   3 & 0.991 & 0.940 &0.941 & 0.957& 0.874 &0.896  & 0.008 & 0.060 &0.065 & 0.042& 0.126&0.116  \\

\hline\hline
\end{tabular}}}
\end{center}
\caption{Simulated coverage probabilities and Type I error rates for different model choices and configurations of $\sigma^{2}$ and $\alpha$ under Scenario 2, for $n_\ell=30,90,150,\ \ell=1,2$.}
\label{tab2}
\end{table}

\subsubsection{Type I error rates}
\label{sssec:t1e}

For the Type I error rate simulations we investigated the two scenarios from Figure~\ref{fig3} for each configuration of
$\alpha = 0.05, 0.1$ and $\sigma^2 = 1, 2, 3$. Further, we set $\delta=\delta_\infty$ in \eqref{H0}.
For a fixed configuration, we generated data according to
\eqref{algorithmus}, fit both models, performed the hypothesis test~(\ref{reject_H}) and counted
the proportion of rejecting the null hypothesis $H$.
Note that due to the choice of $\delta$ both Scenarios 1 and 2 belong to the null hypothesis $H$ defined in (\ref{H0}).
Thus, rejecting $H$ would be a Type I error, i.e. we would erroneously claim similarity of the two dose response curves.

The right side of Table~\ref{tab1} displays the simulated Type I error rates under Scenario 1.
We observe that the simulated Type I error rate is bounded by the nominal significance level $\alpha$ for all configurations investigated here, indicating that
the hypothesis test~(\ref{reject_H}) is indeed a level-$\alpha$ test, even under total sample sizes as small as 30.
Note also that the significance level is actually well exhausted under many configurations.
For small sample sizes and small values of $\delta$ the test becomes conservative, matching the observed
performance of the confidence bounds shown in the left side of Table \ref{tab1}. Again, this conservatism disappears for large sample sizes.

The right side of Table~\ref{tab2} displays the simulated Type I error rates under Scenario 2.
As before, the simulated Type I error rate is bounded by the nominal significance level $\alpha$ under all configurations.
However, we observe that the test is very conservative for small values of $\delta_\infty$,
as already expected from the previously reported results on the coverage probabilities.

\subsubsection{Power}
\label{sssec:pow}

\begin{figure}[h!]
    \begin{center}
      \includegraphics[width=0.5\textwidth]{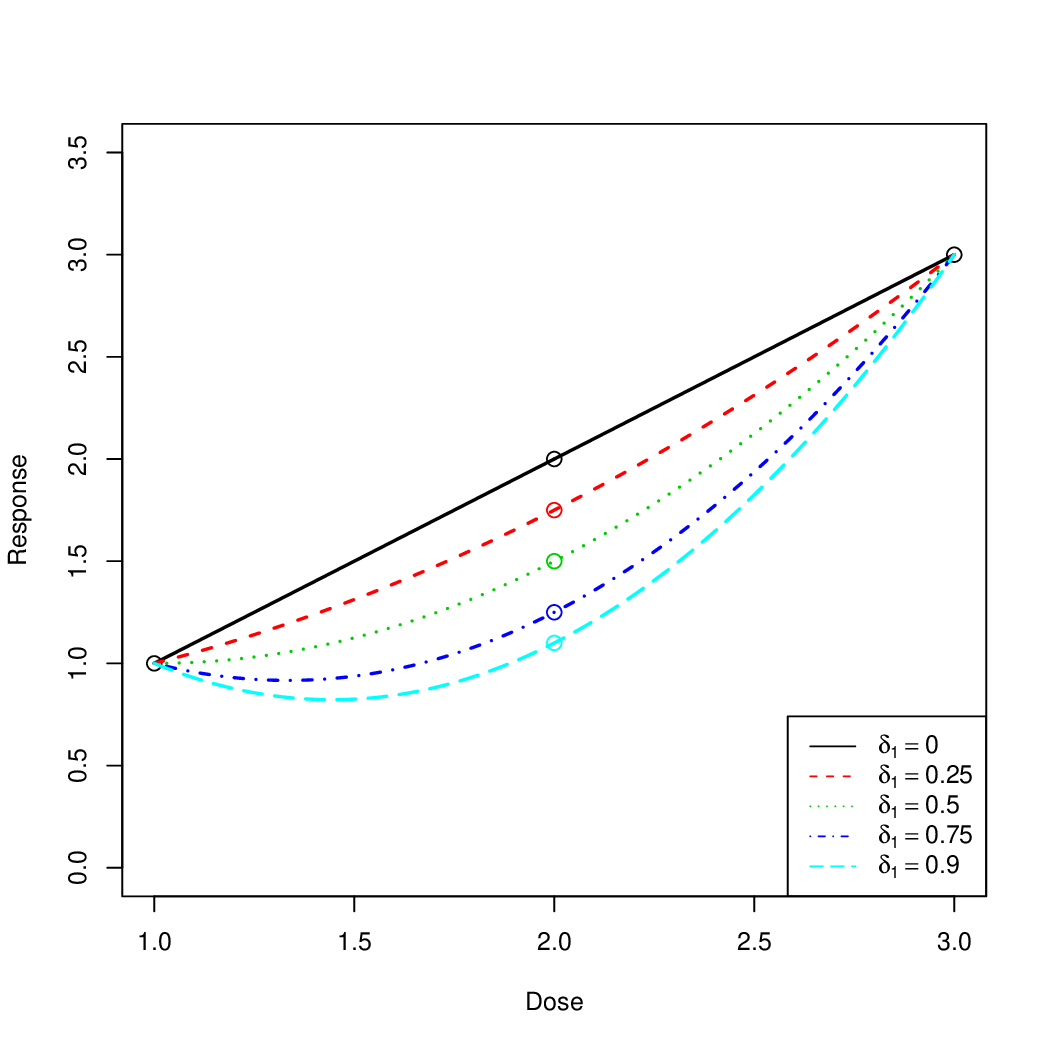}
    \end{center}
\caption{Graphical illustration of Scenario 1 used for the power simulations. Open dots indicate the actual dose levels.} \label{fig4}
\end{figure}

We now consider testing the null hypothesis $H$ in (\ref{H0}) for $\delta = 1$, where in fact the maximum difference is smaller than 1.
We start with the comparison of the models from Scenario 1
for different values of $\delta_1$ under the alternative; see Figure~\ref{fig4}.
The dose levels remain the same as under Scenario 1.
For each configuration of $\sigma^2 = 1, 2, 3$ and $\delta_1 = 0, 0.25, 0.5, 0.75, 0.9$,
we used \eqref{algorithmus} to simulate
$n=10 (30, 50)$ observations under $m_1$ and $m_2$ at each dose level $d_{\ell,i}$, resulting in $n_{\ell} = 30 (90, 150)$, $\ell = 1, 2$.
Table~\ref{tab:power_sim} summarizes the results for $\alpha = 0.05, 0.1$.
The power increases with decreasing values of $\delta_1$.
For large values of $\sigma^2$ the power remains small, even for
$\delta_1 = 0$. In these cases we need larger sample sizes $n_\ell$ in order to
achieve reliable results, as otherwise, due to the large variances, the confidence intervals in \eqref{conf_band} become
too wide and hence the test very conservative.\\

\begin{table}[h!]
\begin{center}
\begin{tabular}{cc|ccccccc} \hline\hline
& & \multicolumn{3}{c}{$\alpha=0.05$} & & \multicolumn{3}{c}{$\alpha=0.1$}\\ \cline{3-5}\cline{7-9}
$\delta_1$ & $\sigma^{2}$ & $n_{\ell}=30$ & $n_{\ell}=90$ & $n_{\ell}=150$ & & $n_{\ell}=30$ & $n_{\ell}=90$ & $n_{\ell}=150$ \\ \hline
  0.00 &   1 & 0.211 & 0.966  & 0.999 & & 0.426 & 0.988  & 0.999 \\
  0.25 &   1 & 0.170 & 0.939  & 0.997 & & 0.377 & 0.974  & 0.999 \\
  0.50 &   1 & 0.102 & 0.731  & 0.917 & & 0.268 & 0.843  & 0.958 \\
  0.75 &   1 & 0.046 & 0.306  & 0.444 & & 0.143 & 0.433  & 0.583 \\
  0.90 &   1 & 0.023 & 0.111  & 0.144 & & 0.074 & 0.195  & 0.245 \\
  0.00 &   2 & 0.002 & 0.544  & 0.911 & & 0.046 & 0.749  & 0.967 \\
  0.25 &   2 & 0.001 & 0.479  & 0.867 & & 0.045 & 0.692  & 0.941 \\
  0.50 &   2 & 0.001 & 0.302  & 0.628 & & 0.030 & 0.500  & 0.770 \\
  0.75 &   2 & 0.000 & 0.119  & 0.247 & & 0.012 & 0.245  & 0.391 \\
  0.90 &   2 & 0.000 & 0.050  & 0.098 & & 0.011 & 0.128  & 0.181 \\
  0.00 &   3 & 0.000 & 0.196  & 0.651 & & 0.007 & 0.434  & 0.822 \\
  0.25 &   3 & 0.000 & 0.162  & 0.576 & & 0.005 & 0.382  & 0.758 \\
  0.50 &   3 & 0.000 & 0.098  & 0.365 & & 0.004 & 0.263  & 0.558 \\
  0.75 &   3 & 0.000 & 0.040  & 0.142 & & 0.002 & 0.128  & 0.276 \\
  0.90 &   3 & 0.000 & 0.021  & 0.050 & & 0.001 & 0.072  & 0.126 \\
\hline\hline
\end{tabular}
\end{center}
\caption{Simulated power for $\delta=1$ and different configurations of $\delta_1$, $\sigma^{2}$, $\alpha$, and $n_{\ell}$ in Scenario 1.}
\label{tab:power_sim}
\end{table}

Regarding Scenario 2, we tested the null hypothesis $H$ in~(\ref{H0}) using $\delta = 1$
and generating data under the models $m_1$, $m_2^1$ and $m_2^2$ defined in~\eqref{scenario2}.
Hence we simulated the performance of the test under the alternative $K$ in~(\ref{H1}) for different choices of $\sigma$ and $\alpha$.
For the sake of brevity we restrict ourselves again to a fixed total sample size of $n_\ell=150$, $\ell=1,2$.
Table \ref{emax2} displays the simulated power. We observe that the test achieves high power, even for larger variances.
However, the power decreases for an increasing true maximum distance between the models and for increasing variances.

\begin{table}[h!]
\begin{center}
\begin{tabular}{ccc|ccccccc} \hline\hline
	& & & \multicolumn{3}{c}{$\alpha=0.05$} & & \multicolumn{3}{c}{$\alpha=0.1$}\\ \cline{4-6}\cline{8-10}
$(m_1,m_2)$ & $\delta_\infty$ & $\sigma^{2}$ &  $n_{\ell}=30$ & $n_{\ell}=90$ & $n_{\ell}=150$ & & $n_{\ell}=30$ & $n_{\ell}=90$ & $n_{\ell}=150$ \\ \hline
 $(m_1,m_1)$ & 0 &            1  & 0.038 & 0.837 & 0.986 & & 0.206&0.930 &0.996  \\
 $(m_1,m_2^1)$ & 0.25 &     1 & 0.036 & 0.770 & 0.980 & & 0.175&0.876 &0.992  \\
$(m_1,m_2^2)$ &  0.5 &     1 & 0.026 & 0.610 & 0.871 & & 0.121 & 0.763 &0.938 \\\hline
$(m_1,m_1)$ & 0 &           2 & 0.001 & 0.257 & 0.719 & & 0.003 &0.517 &0.873  \\
$(m_1,m_2^1)$ & 0.25 &   2 & 0.000 & 0.220 & 0.657 & & 0.005 & 0.493 &0.833  \\
$(m_1,m_2^2)$ &  0.5 &   2 & 0.000 & 0.083 &0.442 & & 0.001 &0.257 &0.655  \\\hline
 $(m_1,m_1)$	& 0 &       3 & 0.000 & 0.023 & 0.350 & & 0.000 & 0.180 &0.622  \\
 $(m_1,m_2^1)$	& 0.25 &  3 & 0.000  & 0.023 &0.286 & & 0.000 & 0.153 &0.553  \\
$(m_1,m_2^2)$	& 0.5 &   3 & 0.000 & 0.010 &0.183 & & 0.000 &0.117 &0.400 \\\hline
\hline\hline
\end{tabular}
\end{center}

\caption{Simulated power for different model choices and configurations of $\sigma^{2}$ and $\alpha$
under Scenario 2, for $\delta=1$ and $n_\ell=30,90,150$, $l=1,2$.}
\label{emax2}
\end{table}

\subsection{Placebo-adjusted modeling}
\label{ssec:adj}

So far we assessed the similarity of two dose response models in terms
of the maximum difference over the dose range under investigation.
Sometimes one might be interested in adjusting for the placebo response, that is,
the treatment effect relative to the placebo response,
before comparing two dose response curves. In this
case one has to modify the results from Section~\ref{ssec:meth1} as follows.
Different to model~(\ref{reg_model}), we consider the placebo-adjusted
responses
\begin{equation*}
Y_{\ell,i,j}=m_\ell\left(\mathbf{\vartheta}_{\ell},d_{\ell,i}\right)-m_\ell\left(\mathbf{\vartheta}_{\ell},0\right)+\epsilon_{\ell,i,j},
\quad j=1, \ldots, n_{\ell,i},\ i=1,\ldots k_\ell,\ \ell=1,2,\ d_{\ell,i}\in \cal{D}.
\end{equation*}
The confidence interval for the maximum absolute difference between the placebo-adjusted curves is then given by
\scriptsize\begin{equation*}
P\left\{ \max_{d\in{\cal D}} |(m_{2}(\mathbf{\vartheta}_{2},d)-m_{2}(\mathbf{\vartheta}_{2},0))-
(m_{1}(\mathbf{\vartheta}_{1},d)-m_{1}(\mathbf{\vartheta}_{1},0))|
\leq \max \big\{ \max_{d\in{\cal D}} U'\left(Y_{1},Y_{2},d\right),-\min_{d\in{\cal D}} L'\left(Y_{1},Y_{2},d\right)\big\} \right\} \geq 1-\alpha,
\end{equation*}\normalsize
where $U'\left(Y_{1},Y_{2},d\right)$ and $L'\left(Y_{1},Y_{2},d\right)$
denote the pointwise confidence bounds for the placebo-adjusted differences derived by the delta method. For example,
\begin{equation*}\label{eq:pointwise}
U'\left(Y_{1},Y_{2},d\right)
= (m_{2}(\hat{\vartheta}_{2},d)-m_{2}(\hat{\vartheta}_{2},0))-
(m_{1}(\hat{\vartheta}_{1},d)-m_{1}(\hat{\vartheta}_{1},0)) + u_{1-\alpha} \hat\rho'(d),
\end{equation*}
where $\hat\rho'(d)$ is calculated for the difference of two placebo-adjusted dose response curves.
Proceeding, the null hypothesis of interest becomes
\begin{eqnarray*}
H':\max_{d\in{\cal D}}\left | (m_{2}(\mathbf{\vartheta}_{2},d)-m_{2}(\mathbf{\vartheta}_{2},0))-
(m_{1}(\mathbf{\vartheta}_{1},d)-m_{1}(\mathbf{\vartheta}_{1},0))\right | \ge\delta
\end{eqnarray*}
and following~(\ref{reject_H}) we reject $H'$ if
\begin{eqnarray}\label{reject_H2}
-\delta < \min_{d\in\cal D} L'\left(Y_{1},Y_{2},d\right)
\quad \mbox{and} \quad  \max_{d\in\cal D} U'\left(Y_{1},Y_{2},d\right) < \delta.
\end{eqnarray}

\begin{figure}[h!]
    \begin{center}
\begin{tabular}{ll}
    $(a)$ & $(b)$ \\
    \includegraphics[width=0.5\textwidth]{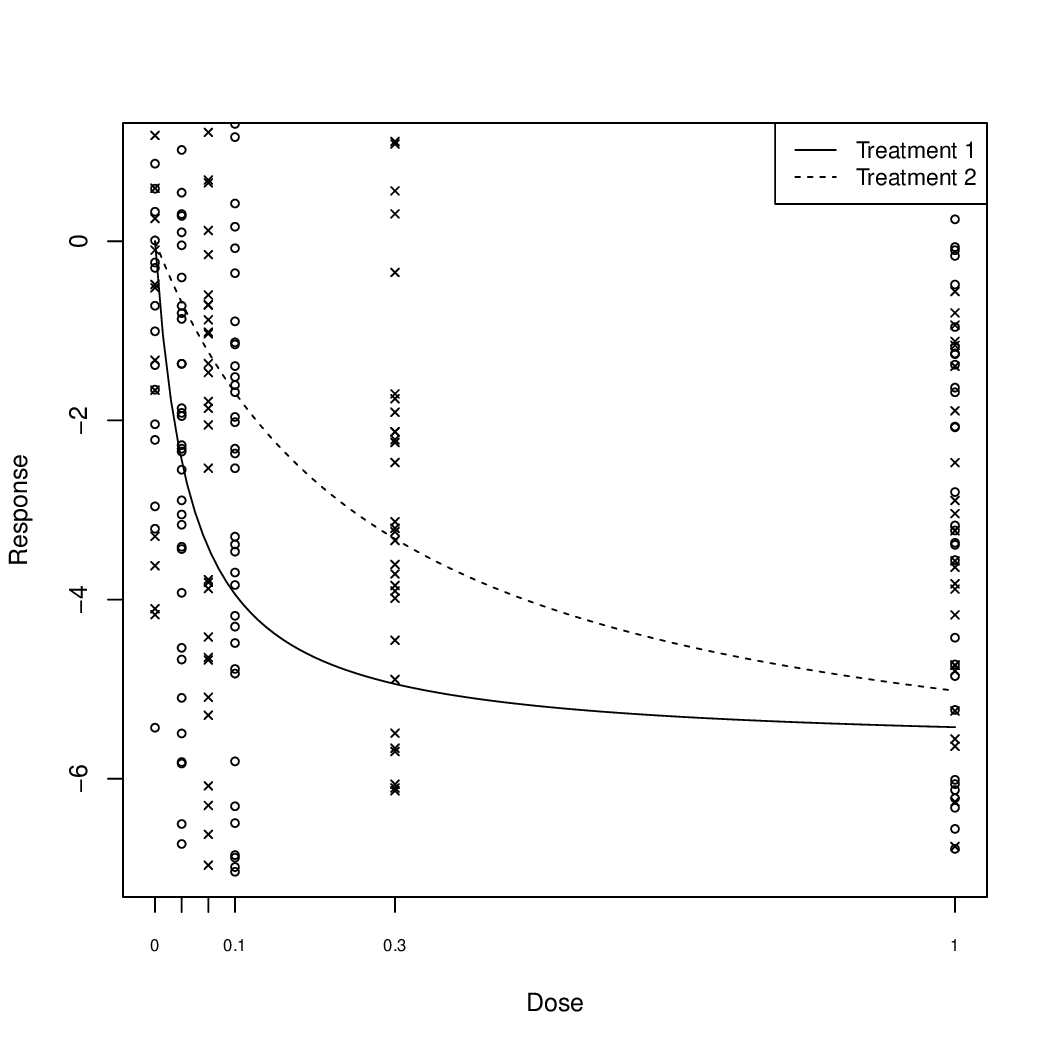} &
    \includegraphics[width=0.5\textwidth]{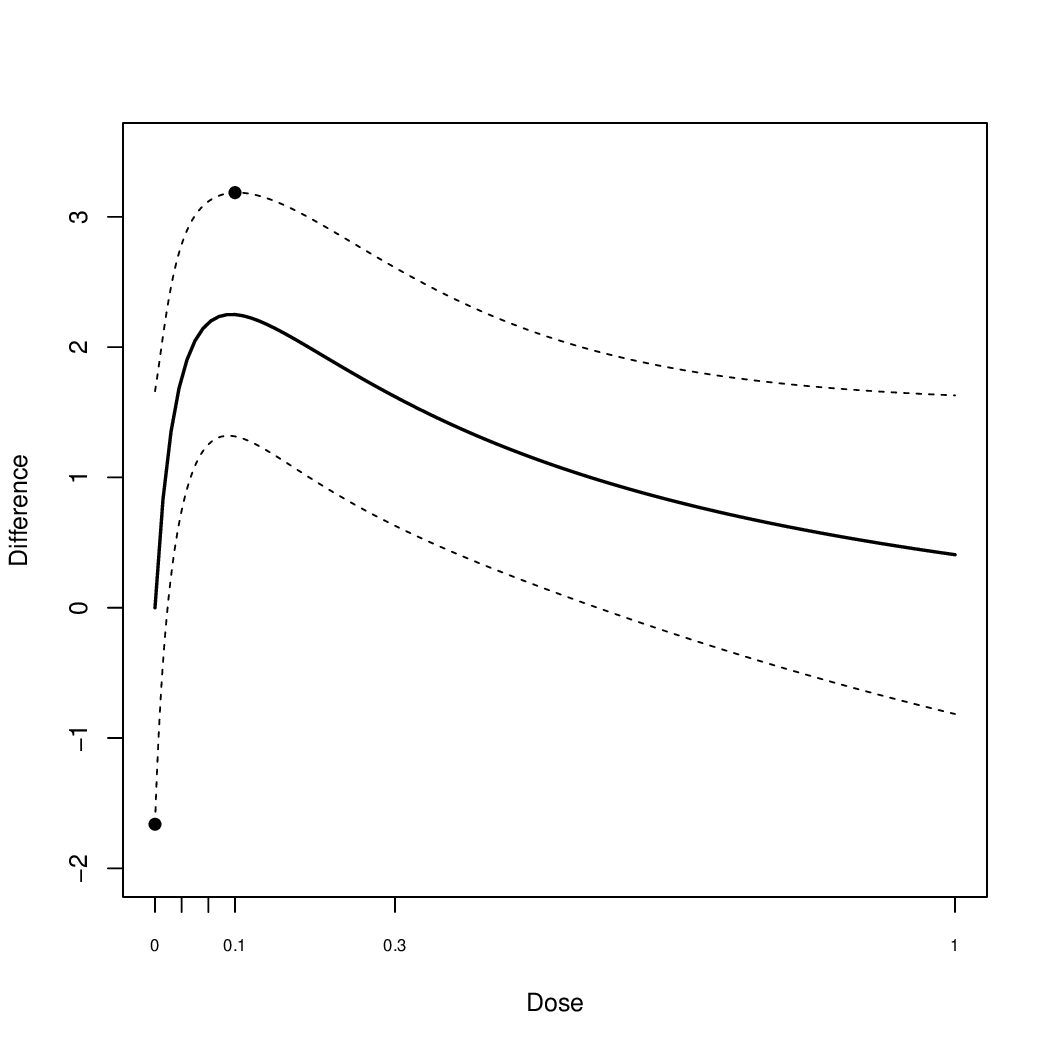}
\end{tabular}
    \end{center}
\caption{Placebo-adjusted plots for the weight loss case study.
$(a)$ The placebo-adjusted Emax model fit $m_1$ ($m_2$) for the o.d. (b.i.d.) regimen is given by the solid (dashed) line with observations marked by ``x'' (``o'').
$(b)$ Mean difference curve with associated pointwise 90\% confidence bounds. Bold dots denote the maximum upper and minimum lower confidence bound over ${\cal D} = [0,1]$.}
\label{fig5}
\end{figure}

To illustrate this methodology, we revisit the weight loss case study from Section~\ref{ssec:case1}.
The individual model fits remain the same, i.e. $m_1(\vartheta_1, d) = 0.55 - 5.66 \frac{d}{6.55 + d}$ for the o.d. regimen
and $m_2(\vartheta_2, d) = -0.54 - 6.42\frac{d}{41.99 + d}$ for the b.i.d. regimen.
Figure~\ref{fig5}$a$ displays the placebo-adjusted model fits $m_1(\hat{\vartheta}_{1}, d)-m_1(\hat{\vartheta}_{1}, 0)$
and $m_2(\hat{\vartheta}_{2}, d)-m_2(\hat{\vartheta}_{2}, 0)$, $d \in [0, 1]$,
together with the individual observations, where only the range $[-7, 1]$ is displayed on the vertical axis for better readability.
Figure~\ref{fig5}$b$ displays
the difference $(m_2(\hat{\vartheta}_{2}, d)-m_2(\hat{\vartheta}_{2}, 0)) -
(m_1(\hat{\vartheta}_{1}, d)-m_1(\hat{\vartheta}_{1}, 0))$
together with the associated 90\% pointwise confidence intervals for each dose
$d \in [0, 1]$.
In this example, the estimated placebo effects from the original fits were slightly different to 0.
Thus, the placebo-adjusted difference curve and its confidence bounds differ slightly from the previous results in Section $2.2$; see Figure~\ref{fig2}.
The maximum upper confidence bound for $\alpha=0.1$ is
$\max_{d\in\cal D}U'\left(Y_{1},Y_{2},d\right)=3.186$,
again observed at dose $d=0.1$, and the minimum lower confidence bound is
$\min_{d\in\cal D} L'\left(Y_{1},Y_{2},d\right)=-1.661$ at dose $d=0$.
That is, the maximum placebo-adjusted difference between the two regimens over the dose range
$\mathcal{D}=\left[0,1\right]$ lies between $-1.661$ and $3.186$. Therefore, similarity  of the placebo-adjusted dose response curves
can be claimed according to (\ref{reject_H2}) as long as $\delta$ is larger than $3.186$.

\section{Assessing the similarity of two target doses}
\label{sec:td}

This section focuses on assessing the similarity of two target doses.
We consider the difference between the minimum effective doses ($MED$s) of two dose response curves from two non-overlapping subgroups.
We derive confidence intervals and statistical tests to decide at a given level $\alpha$ whether the absolute difference of two $MED$s is smaller than a prespecified margin $\eta$.
Furthermore, we illustrate the proposed methodology by revisiting the case study from~\ref{ssec:case1} and investigate its operating characteristics.

\subsection{Methodology}
\label{ssec:meth2}

Following \cite{Ruberg95a}, the $MED$ is defined as the smallest dose that produces a clinically relevant response $\Delta$ on top of the placebo effect (i.e. at dose $d=0$).
That is,
\begin{equation}
    MED_\ell=MED_\ell(\vartheta_\ell)=\inf_{d\in\cal D}\left\{m_\ell(\vartheta_\ell,0)<m_\ell(\vartheta_\ell,d)-\Delta\right\},\ \ell=1,2. \label{MED0}
\end{equation}
From now on we assume strict monotonicity of the dose response curves $m_\ell$ such that \eqref{MED0} becomes
\begin{equation*}
    MED_\ell=MED_\ell(\vartheta_\ell)=m_\ell^{-1}(\vartheta_\ell,m_\ell(\vartheta_\ell,0)+\Delta), \ell=1,2,
\end{equation*}
where the inverse is calculated with respect to $d$ for fixed model parameters $\vartheta_1$ and $\vartheta_2$.
Estimates for the $MED$ are then given by
\begin{equation*}
    \widehat{MED}_\ell=m_\ell^{-1}(\hat{\vartheta}_\ell,m_\ell(\hat\vartheta_\ell,0)+\Delta), \ell=1,2,
\end{equation*}
where $\hat{\vartheta}_1$ and $\hat{\vartheta}_2$ are the non-linear least squares estimators for the true parameters.
Due to the asymptotic normality of the estimates $\hat{\vartheta}_1$ and $\hat{\vartheta}_2$, the estimated difference of the $MED$s is approximately normal distributed \cite{DetteBretz2007}.
To be more precise, the delta method \cite{Serfling1980} gives
\begin{equation}\widehat{MED}_1-\widehat{MED}_2-(MED_1-MED_2)\approx\mathcal{N}(0,\tau^2),\label{MEDverteilung}\end{equation}
for
\begin{equation*}
    \tau^2=\left(\tfrac{\partial}{\partial\vartheta_1} m_1^{-1}(\vartheta_1,\Delta_1)\right)^T\tfrac{\sigma_1^2}{n_1}\Sigma_1^{-1}\tfrac{\partial}{\partial\vartheta_1} m_1^{-1}(\vartheta_1,\Delta_1)+\left(\tfrac{\partial}{\partial\vartheta_2} m_2^{-1}(\vartheta_2,\Delta_2)\right)^T\tfrac{\sigma_2^2}{n_2}\Sigma_2^{-1}\tfrac{\partial}{\partial\vartheta_2} m_2^{-1}(\vartheta_2,\Delta_2) \label{tau}
\end{equation*}
and $\Delta_\ell=m_\ell(\vartheta_\ell,0)+\Delta,\ \ell=1,2$.
\HD{The variance $\tau^2$ can be estimated by replacing $\vartheta_\ell$ and $\Sigma_\ell$ by their  estimates $\hat \vartheta_\ell$
and $\hat \Sigma_\ell$, $\ell=1,2$; see Section~\ref{ssec:meth1}. The corresponding estimator is denoted by $\hat \tau^2$.
It then follows from \eqref{MEDverteilung} that
\small
\begin{equation}
    P\left\{MED_1-MED_2\in \left[\widehat{MED}_1-\widehat{MED}_2-u_{1-\alpha/2}\hat{\tau},\widehat{MED}_1-\widehat{MED}_2+u_{1-\alpha/2}\hat{\tau}\right]\right\}
   \stackrel{n_1,n_2\rightarrow \infty} {\longrightarrow}1-\alpha,
    \label{conf_MED}
\end{equation}
\normalsize
and an asymptotic $(1- \alpha )$-confidence interval for the difference of the $MED$s is given by
$$\big [\widehat{MED}_1-\widehat{MED}_2-u_{1-\alpha/2}\hat{\tau},\widehat{MED}_1-\widehat{MED}_2+u_{1-\alpha/2}\hat{\tau}\big].$$
In order to derive a test for similarity of two target doses we consider the problem of testing
\begin{equation}
    H''\colon |MED_1-MED_2| \geq \eta \qquad \text{against} \qquad K''\colon |MED_1-MED_2|< \eta.\label{hypothesesMED}
\end{equation}
In Appendix~B we show that rejecting $H''$ if
\begin{equation}
    |\widehat{MED}_1-\widehat{MED}_2|<c \label{testMED},
\end{equation}
gives an asymptotic (uniformly most powerful) level $\alpha$ test, where $c$ is the unique solution of the equation
\begin{equation}
    \alpha=\Phi\left(\frac{c-\eta}{\hat{\tau}}\right)-\Phi\left(\frac{-c-\eta}{\hat{\tau}}\right).\label{constant_c}
\end{equation}
Note that \eqref{constant_c} can easily be solved by using Newton's algorithm \cite{Deuflhard}.}

\subsection{Case study revisited}
\label{ssec:case2}

\begin{figure}[h!]
    \begin{center}
\begin{tabular}{ll}
    $(a)$ & $(b)$ \\
    \includegraphics[width=0.5\textwidth]{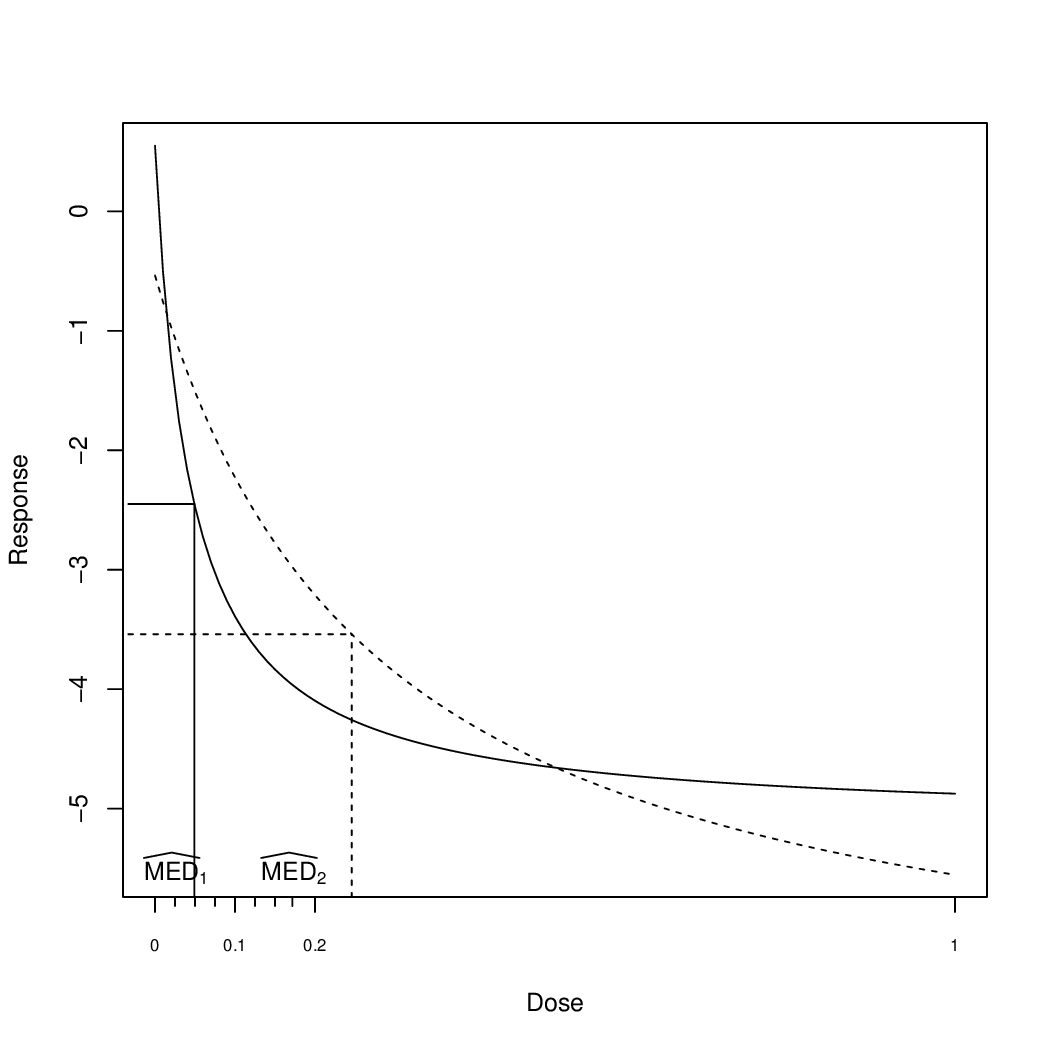} &
    \includegraphics[width=0.5\textwidth]{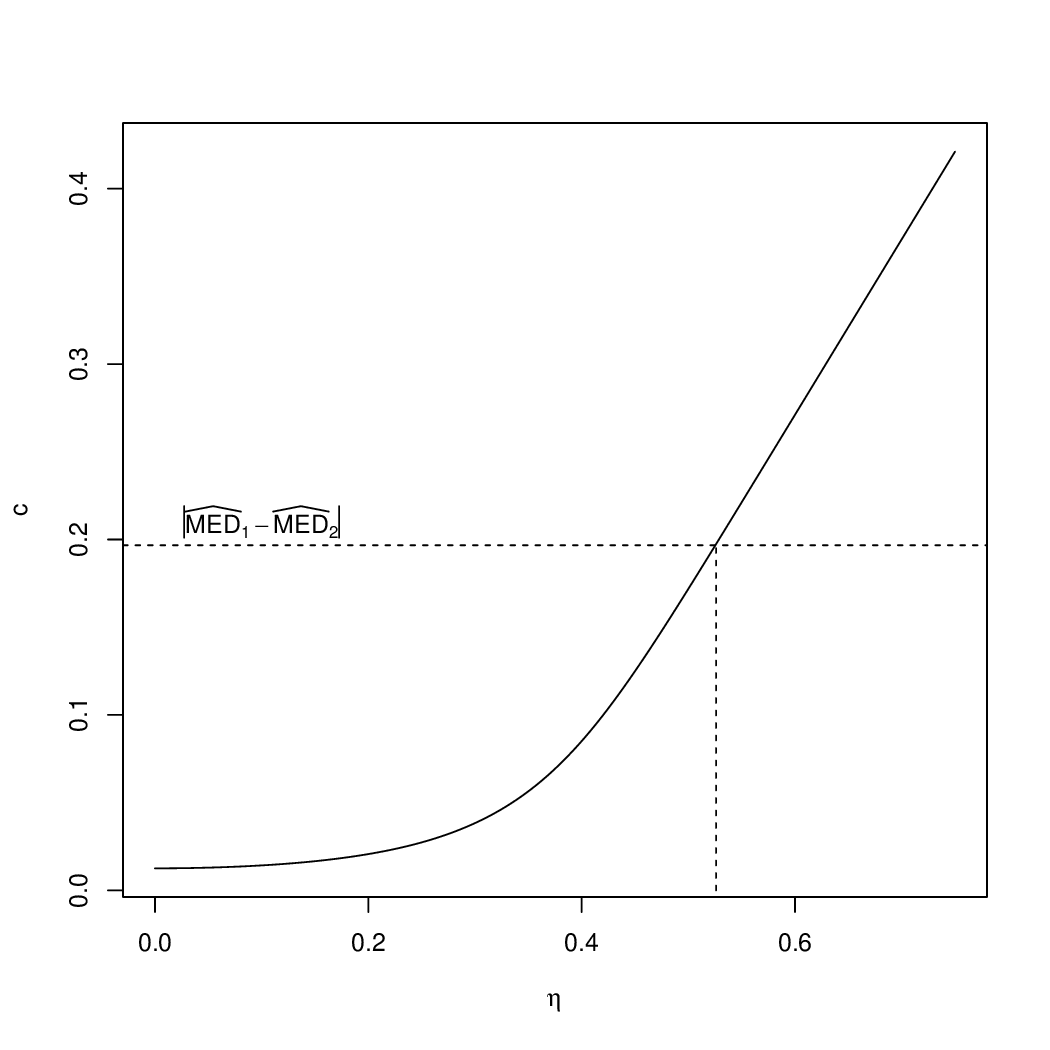}
\end{tabular}
    \end{center}
\caption{Plots for the revisited weight loss case study.
$(a)$ The fitted Emax model $m_1$ ($m_2$) for the o.d. (b.i.d.) regimen is given by the solid (dashed) line, together with the estimated $MED$s for $\Delta=-3$.
$(b)$ Plot of the unique solution  $c$ of equation \eqref{constant_c} as a function of $\eta$.
The dashed lines indicate the absolute difference of the $MED$ estimates and the minimum choice of $\eta$ in order to claim similarity for $\alpha=0.05$.}
\label{MED_fig_casestudy}
\end{figure}

To illustrate the methodology in the previous subsection, we revisit the weight loss case study from Section~\ref{ssec:case1}.
Recall the individual model fits $m_1(\hat{\vartheta}_1, d) =0.55 - 5.66 \frac{d}{6.55 + d}$ for the o.d. regimen
and $m_2(\hat{\vartheta}_2, d) =  -0.54 - 6.42\frac{d}{41.99 + d}$ for the b.i.d. regimen.
We chose a clinically relevant difference of $\Delta=-3$. That is, a weight loss of $3\%$ compared to the placebo response
is assumed to be a clinically relevant effect on top of the placebo response at dose $d=0$.
Therefore, $\widehat{MED}_1=m_1^{-1}(\hat\vartheta_1,0.55-3)=0.049$, $\widehat{MED}_2=m_2^{-1}(\hat\vartheta_2,-0.54-3)=0.246$ and $\widehat{MED}_1-\widehat{MED}_2=-0.196$.
Figure~\ref{MED_fig_casestudy}$(a)$ displays the model fits $m_\ell(\hat{\vartheta}_\ell, d)$, together with the estimates $\widehat{MED}_\ell$, $\ell=1,2$.

The $1-\alpha$ confidence interval for the true difference $MED_1-MED_2$ is then given by $\left[-0.197-u_{1-\alpha/2}0.199,-0.197+u_{1-\alpha/2}0.199\right]$.
For example, $MED_1-MED_2 \in \left[-0.589,0.195\right]$ for $\alpha=0.05$ and $MED_1-MED_2 \in \left[-0.526,0.132\right]$ for $\alpha=0.1$.
Applying the test in \eqref{testMED} for $\alpha=0.05$ allows us to claim similarity of the two $MED$s whenever $\eta>0.526$ because of
$c>0.197=|\widehat{MED}_1-\widehat{MED}_2|$ in \eqref{constant_c}.
Figure \ref{MED_fig_casestudy}$(b)$ displays the value of $c$ as a function of $\eta$.
For $\alpha=0.1$ we obtain by similar calculations that $\eta$ has to be larger than $0.453$ in order to claim similarity.

\subsection{Simulations}
\label{ssec:sim2}

We now report the results of a simulation study to investigate the operating characteristics of the method described in Section~\ref{ssec:meth2}.
Adapting the data generation algorithm from Section~\ref{ssec:sim1}, we investigated the coverage probabilities of the confidence intervals in \eqref{conf_MED} as well as the Type I error rates and
power of the test~\eqref{testMED} for different scenarios.
All results were obtained using $10,000$ simulation runs. Again we refer to \cite{Bretz2016} for the complete simulations results.

\begin{figure}[h!]
    \begin{center}
\begin{tabular}{ll}
    $(a)$ & $(b)$ \\
    \includegraphics[width=0.5\textwidth]{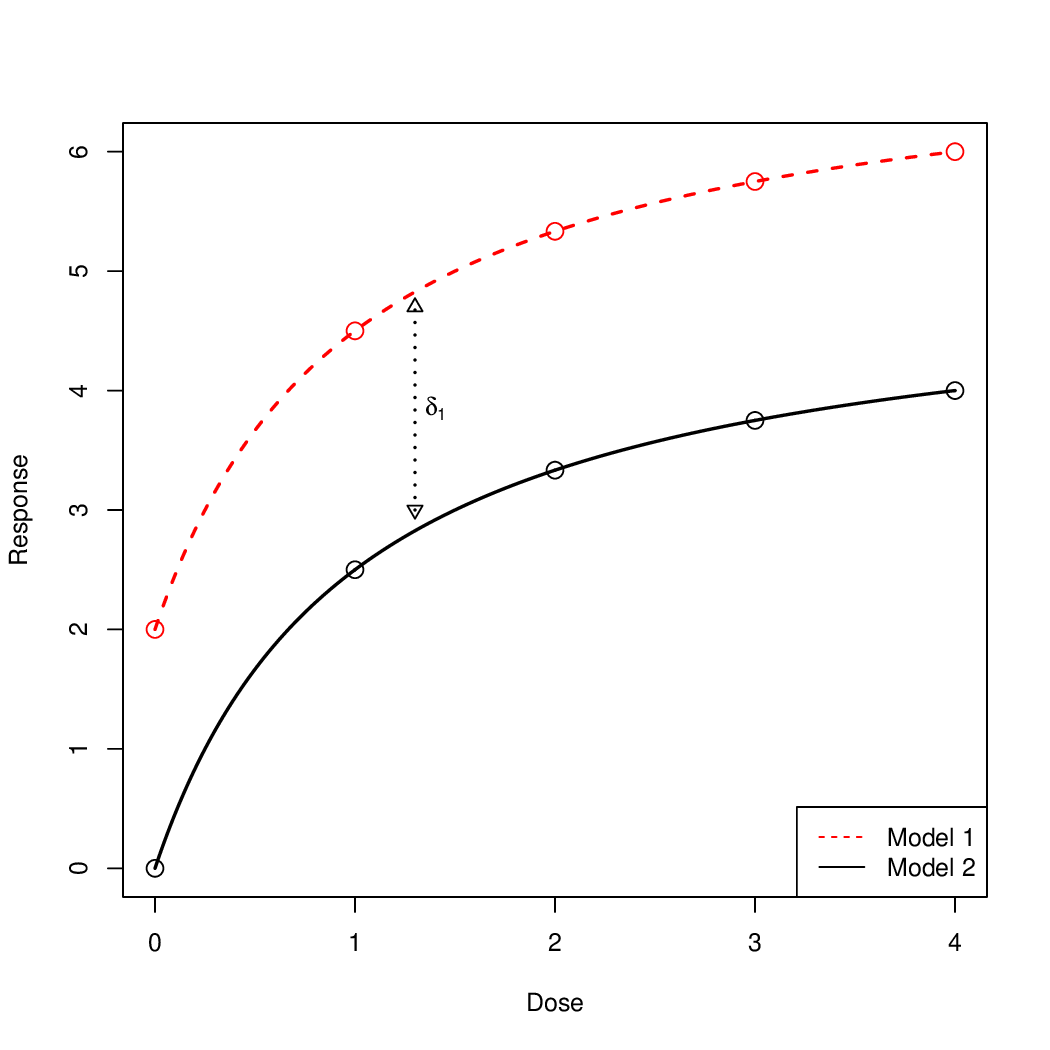} &
    \includegraphics[width=0.5\textwidth]{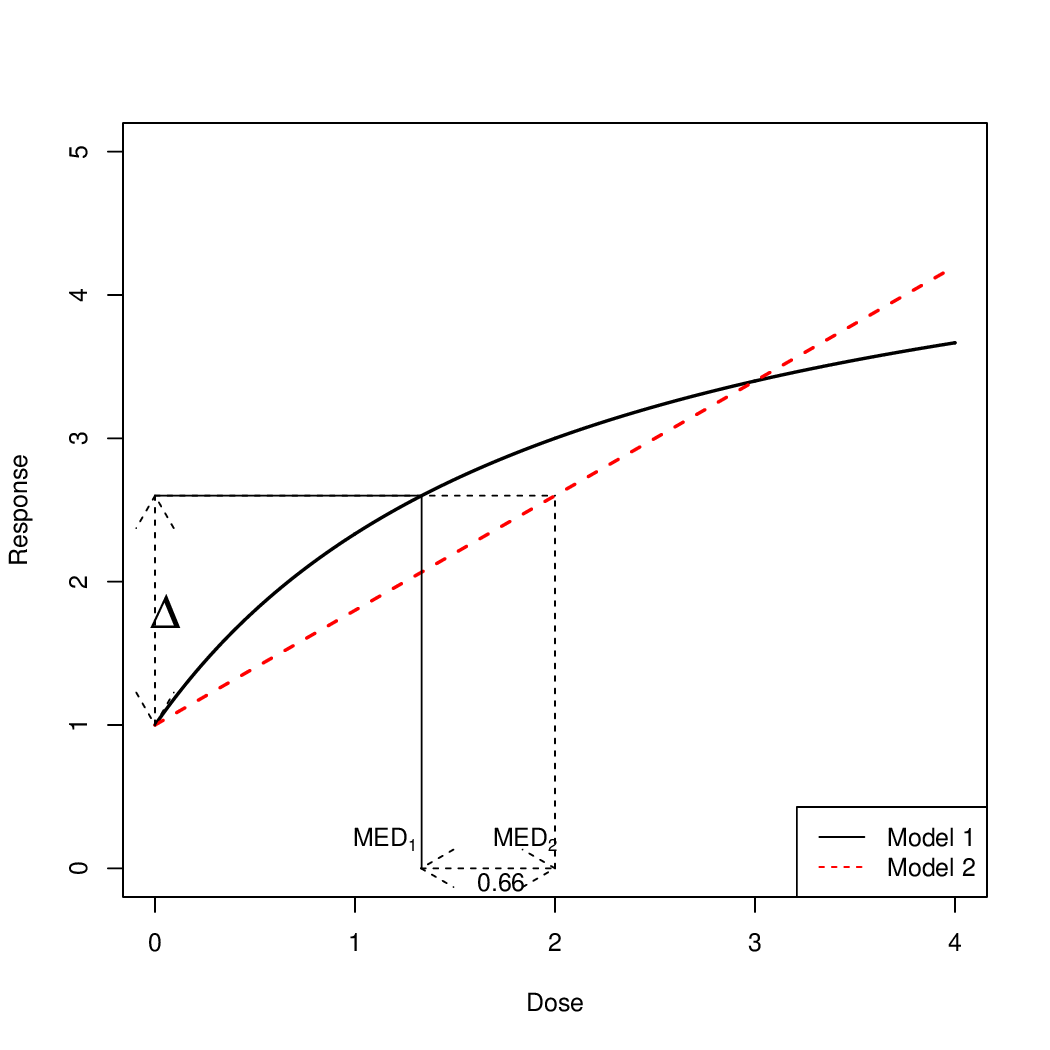}
\end{tabular}
    \end{center}
\caption{Graphical illustration of Scenarios 3 and 4 used for the simulations. $(a)$ displays the shifted Emax models with $\delta_1=2$. $(b)$ displays the curves for Scenario 4, together with the $MED$s corresponding to $\Delta=1.6$.}
\label{MED_fig}
\end{figure}

\subsubsection{Coverage probabilities}
\label{sssec:cp2}

\paragraph{Scenario 3}
We start with the comparison of two shifted Emax models $m_1(d,\vartheta_1)=\delta_1+5d/(1+d)$ and $m_2(d,\vartheta_2)=5d/(1+d)$
over $\mathcal{D}=[0,4]$, with identical dose levels $d_{\ell,i}=i-1, i=1, \ldots, 5$ for both regression models $\ell=1,2$; see Figure~\ref{MED_fig}$a$.
Because the models are shifted by the constant $\delta_1$, the true difference $MED_1-MED_2=0$ regardless of the value for $\Delta$.
For each configuration of $\sigma^2 = 1, 2$ and $\delta_1=1, 2, 3$ we used \eqref{algorithmus} to simulate
$n_{\ell,i}=6(30)$ observations at each dose level $d_{\ell,i}$, resulting in $n_{\ell} = 30(150)$, $\ell = 1, 2$.

The left side of Table~\ref{tab1MED} displays the coverage probabilities for $\alpha = 0.05, 0.1$.
We observe that the coverage probability is at least $1-\alpha$ under all configurations.
The confidence intervals are more accurate for larger sample sizes and smaller variances, which confirms the asymptotic result from \eqref{conf_MED}.
Furthermore, the simulated differences between the $MED$ estimates are very close to the true difference under all configurations (results not shown here).

%

\begin{table}[h!]
	\tiny
\begin{center}
\makebox[\linewidth]{ 
\begin{tabular}{cc|ccc|ccc|ccc|ccc} \hline\hline
& & \multicolumn{6}{c}{Coverage probabilities} \vline & \multicolumn{6}{c}{Type I error rates}  \\
\cline{3-14}
& & \multicolumn{3}{c}{$\alpha=0.05$} &  \multicolumn{3}{c}{$\alpha=0.1$} \vline & \multicolumn{3}{c}{$\alpha=0.05$} & \multicolumn{3}{c}{$\alpha=0.1$}\\ \cline{3-14}
$\delta_1$ & $\sigma^{2}$ & $n_{\ell}=30$ & $n_{\ell}=90$ & $n_{\ell}=150$ & $n_{\ell}=30$ & $n_{\ell}=90$ & $n_{\ell}=150$ & $n_{\ell}=30$ &$n_{\ell}=90$ & $n_{\ell}=150$  & $n_{\ell}=30$& $n_{\ell}=90$ & $n_{\ell}=150$ \\ \hline
1 & 1 & 0.979 & 0.948 &0.959 & 0.941 &  0.926 &0.907 & 0.050 & 0.048 &0.050  & 0.103 & 0.110 &0.103\\
2 & 1 & 0.982 & 0.962 &0.958 & 0.945 & 0.917 &0.909 & 0.053 & 0.051 &0.048 & 0.105 & 0.102 &0.105\\
3 & 1 & 0.980 & 0.972 &0.961 & 0.946 & 0.922  &0.908 & 0.053 & 0.052 &0.052 & 0.099 & 0.100 &0.105\\
1 & 2 & 0.996 & 0.968 &0.967 & 0.977 & 0.948 &0.917 & 0.049 & 0.049 &0.051  & 0.104 & 0.104 &0.101\\
2 & 2 & 0.996 & 0.969 &0.968 & 0.978 & 0.959 &0.922 & 0.052 & 0.050 &0.049 & 0.103 & 0.100 &0.101\\
3 & 2 & 0.995 & 0.977 &0.966 & 0.976 & 0.923  &0.916 & 0.045 & 0.048 &0.049 & 0.100 & 0.098 &0.099\\
1 & 3 & 0.999 & 0.978 &0.977 & 0.979 & 0.941 &0.927 & 0.050 & 0.062 &0.051  & 0.104 & 0.110 &0.101\\
2 & 3 & 0.998 & 0.971 &0.969 & 0.979 & 0.952 &0.925 & 0.058 & 0.058 &0.049 & 0.103 & 0.102 &0.111\\
3 & 3 & 0.995 & 0.981 &0.967 & 0.978 & 0.923 &0.918 & 0.044 & 0.049 &0.049 & 0.100 & 0.092 &0.088\\
\hline\hline
\end{tabular}}
\end{center}
\caption{Simulated coverage probabilities and Type I error rates for different configurations of $\delta_1$, $\sigma^{2}$, $\alpha$, and $n_{\ell}$ under Scenario 3.}
\label{tab1MED}
\end{table}

\paragraph{Scenario 4}
We now consider the comparison of the Emax model $m_1(d,\vartheta_1)=1+4d/(2+d)$ with the linear model $m_2(d,\vartheta_2)=1+0.8d$ for the same set of doses as in Scenario 3.
Note that the responses at doses $d=0$ and $d=3$ are the same in both models; see Figure \ref{MED_fig}$b$.
For each configuration of $\sigma^2 = 1, 2, 3$ and $\Delta = 0.8, 1.6, 2.4$, we used again \eqref{algorithmus} to simulate $n_{\ell,i}=6(30)$ observations at each dose level $d_{\ell,i}$, resulting in $n_{\ell} = 30(150)$, $\ell = 1, 2$.

The left side of Table~\ref{tab3MED} displays the coverage probabilities for $\alpha = 0.05, 0.1$.
As before, asymptotically the coverage probability is at least $1-\alpha$ under all configurations investigated here,
except for small sample sizes and $\Delta=2.4$ (in which case the $MED$s coincide).
This is a direct consequence of the definition of the $MED$.
Inverting an Emax model $m(\vartheta,d) = y = \vartheta_{1} + \vartheta_{2}d/(\vartheta_{3}+d)$
gives $m_1^{-1}(\vartheta,y)=\vartheta_{3}(y-\vartheta_{1})/(\vartheta_{1}+\vartheta_{3}-y)$.
Therefore higher values of $\Delta$ result in being closer to the pole of $m^{-1}$, which is at $\vartheta_{1}+\vartheta_{2}=5$ in this case.
However, further simulations show that the results get better for larger sample sizes and the coverage probabilities converge quickly to their nominal values.
Finally, the simulated differences between the $MED$ estimates are very close to the true difference under all configurations, except in the case where
 the $MED$s coincide (i.e. $\Delta=2.4$; results not shown here).

\begin{table}[h!]
	\tiny
\begin{center}
\makebox[\linewidth]{ 
	\begin{tabular}{cc|ccc|ccc|ccc|ccc} \hline\hline
		& & \multicolumn{6}{c}{Coverage probabilities} \vline & \multicolumn{6}{c}{Type I error rates}  \\
		\cline{3-14}
		& & \multicolumn{3}{c}{$\alpha=0.05$} &  \multicolumn{3}{c}{$\alpha=0.1$} \vline & \multicolumn{3}{c}{$\alpha=0.05$} & \multicolumn{3}{c}{$\alpha=0.1$}\\ \cline{3-14}
		$\Delta$ & $\sigma^{2}$ & $n_{\ell}=30$ & $n_{\ell}=90$ & $n_{\ell}=150$ & $n_{\ell}=30$ & $n_{\ell}=90$ & $n_{\ell}=150$ & $n_{\ell}=30$ &$n_{\ell}=90$ & $n_{\ell}=150$  & $n_{\ell}=30$& $n_{\ell}=90$ & $n_{\ell}=150$ \\ \hline
0.8 & 1 & 0.964 & 0.960 & 0.965 & 0.929 & 0.914 & 0.908 & 0.036& 0.025& 0.025  & 0.077 &0.059 & 0.068\\
1.6 & 1 & 0.953 & 0.951 & 0.946 & 0.922 & 0.913 & 0.903 & 0.051& 0.048 & 0.040  & 0.098 &0.088 & 0.087\\
2.4 & 1 & 0.920 & 0.932 & 0.949 & 0.877 & 0.900 & 0.916 & 0.069& 0.055 & 0.057  & 0.137 &0.121  &0.116\\
0.8 & 2 & 0.989 &0.953 & 0.968 & 0.960 & 0.924 & 0.928 & 0.050& 0.038 & 0.026  & 0.101 & 0.063 &0.061\\
1.6 & 2 & 0.967 &0.964 & 0.956 & 0.936 & 0.907 & 0.913 & 0.053& 0.046 & 0.045  & 0.111 & 0.087 &0.088\\
2.4 & 2 & 0.918 &0.919 & 0.932 & 0.870 &0.888 & 0.901 & 0.069& 0.054 & 0.060  & 0.143 & 0.125 &0.124\\
0.8 & 3 & 0.969 &0.959 & 0.969 & 0.925 &0.921& 0.945 & 0.055& 0.039 & 0.029  & 0.100 & 0.092 &0.074\\
1.6 & 3 & 0.989&0.971 & 0.964 & 0.913 & 0.902 & 0.919 & 0.044& 0.046 & 0.043  & 0.113 & 0.096 &0.098\\
2.4 & 3 & 0.915 & 0.922 & 0.912 & 0.860 &0.879& 0.892 & 0.070& 0.064 & 0.061  & 0.145 & 0.123 &0.119\\
\hline\hline
\end{tabular}}
\end{center}
\caption{Simulated coverage probabilities and Type I error rates for different configurations of $\Delta$, $\sigma^{2}$, $\alpha$, and $n_{\ell}$ under Scenario 4.}
\label{tab3MED}
\end{table}
%

\subsubsection{Type I error rates}
\label{sssec:t1e2}

For the Type I error rate simulations we investigated the two scenarios from Figure~\ref{MED_fig}.
We start with Scenario 3. Because $|MED_1-MED_2|=0$ for all values of $\Delta$, we chose $\eta=0$.
For a fixed configuration of parameters, we generated data according to \eqref{algorithmus}, fit both models,
performed the hypothesis test~(\ref{testMED}) and counted the proportion of rejecting the null hypothesis $H''$.
The right side of Table~\ref{tab1MED} displays the simulated Type I error rates under Scenario 3.
We observe that the simulated Type I error rate is well exhausted at the nominal significance level $\alpha$ for all configurations investigated here, indicating that
the hypothesis test~(\ref{testMED}) is indeed a level-$\alpha$ test, even under total sample sizes as small as 30.

The right side of Table~\ref{tab3MED} displays the simulated Type I error rates under Scenario 4.
As before, the simulated Type I error rate is bounded by the nominal significance level $\alpha$ under almost all configurations.
The test can be liberal for small sample sizes and large values of $\Delta$, matching the observed
performance of the confidence bounds shown in the left side of Table \ref{tab3MED}. Again, this size inflation disappears for large sample sizes.



\subsubsection{Power}
\label{sssec:pow2}

For the power simulations we again considered the two scenarios from Figure~\ref{MED_fig} and start with Scenario 3.
Because $|MED_1-MED_2|=0$ for all values of $\Delta$, the power of the test depends only on the given threshold $\eta$.
For the concrete simulations, we set $\Delta=1$ and used $\delta_1=1$ for convenience.
For each configuration of $\sigma^2 = 1, 2, 3$ and $\eta = 0.1, 0.2, 0.5, 1$, we used \eqref{algorithmus} to simulate $n=10 (30, 50)$ observations
under $m_1$ and $m_2$ at each dose level $d_{\ell,i}$, resulting in $n_{\ell} = 30 (90, 150)$, $\ell = 1, 2$.
All configurations belong to the alternative in \eqref{hypothesesMED}.
Table~\ref{tab2MED} summarizes the results for $\alpha = 0.05, 0.1$.
The power increases with increasing values of $\eta$.
The power decreases for larger values of $\sigma^2$, especially for small values of $\eta$.
In these cases we need larger sample sizes $n_\ell$ in order to achieve reliable results.

\begin{table}[h!]
	\begin{center}
		\begin{tabular}{cc|ccccccc} \hline\hline
			& & \multicolumn{3}{c}{$\alpha=0.05$} & & \multicolumn{3}{c}{$\alpha=0.1$}\\ \cline{3-5}\cline{7-9}
			$\eta$ & $\sigma^{2}$ & $n_{\ell}=30$ & $n_{\ell}=90$ & $n_{\ell}=150$ & & $n_{\ell}=30$ & $n_{\ell}=90$ & $n_{\ell}=150$ \\ \hline
			1 & 1    & 0.979& 1.000 & 1.000 & & 0.989&1.000 & 1.000\\
			0.5 & 1  & 0.679& 0.988 & 0.999 & & 0.783&0.995 & 1.000\\
			0.2 & 1  & 0.116& 0.364 & 0.641 & & 0.226&0.543 & 0.784\\
			0.1 & 1  & 0.061& 0.086 & 0.123 & & 0.123&0.179 & 0.232\\
			1 & 2    & 0.823& 0.997 & 1.000 & & 0.893&0.999 & 1.000\\
			0.5 & 2  & 0.400& 0.853 & 0.975 & & 0.524&0.915 & 0.987\\
			0.2 & 2  & 0.078& 0.167 & 0.283 & & 0.163&0.288 & 0.462\\
			0.1 & 2  & 0.055& 0.066 & 0.077 & & 0.109&0.132 & 0.156\\
            1 &   3   & 0.703& 0.985 & 1.000 & & 0.794&0.989 & 1.000\\
			0.5 & 3 & 0.265& 0.691 & 0.897 & & 0.401&0.788& 0.936\\
			0.2 & 3  & 0.075& 0.112 & 0.168 & & 0.145&0.214 & 0.311\\
			0.1 & 3  & 0.047& 0.050 & 0.068 & & 0.116&0.132 & 0.130\\
			\hline\hline
		\end{tabular}
	\end{center}
	\caption{Simulated power for different configurations of $\eta$, $\sigma^{2}$, $\alpha$, and $n_{\ell}$ in Scenario 3.}
	\label{tab2MED}
\end{table}

For the final set of simulations, we revisit Scenario 4 and investigate the power for different values of $\sigma^2$ and $\Delta$.
We set $\eta=0.8$ and $n_\ell = 30,150$ for $\ell=1,2$ and summarize the results in Table \ref{tab6MED}.
In alignment with all former results, the performance of the test is worse in case of $\Delta=2.4$ due to the already mentioned numerical problems when calculating the $MED$s.
In general, the power increases with  increasing sample sizes and  decreasing variances under all observed configurations.
The power converges  to $1$ for $n_1,n_2 \to \infty$.

\begin{table}[h!]
	\begin{center}
	\begin{tabular}{cc|ccccccc} \hline\hline
		& & \multicolumn{3}{c}{$\alpha=0.05$} & & \multicolumn{3}{c}{$\alpha=0.1$}\\ \cline{3-5}\cline{7-9}
		$\Delta$ & $\sigma^{2}$ & $n_{\ell}=30$ & $n_{\ell}=90$ & $n_{\ell}=150$ & & $n_{\ell}=30$ & $n_{\ell}=90$ & $n_{\ell}=150$ \\ \hline
0.4 & 1 & 0.914 &1.000 & 1.000 & & 0.958 & 1.000 & 1.000\\
0.8 & 1 & 0.116 &0.404 & 0.625 & & 0.261 &0.609 & 0.778\\
1.6 & 1 & 0.057 &0.080 & 0.080 & & 0.118 &0.129 & 0.152\\
2.4 & 1 & 0.090 &0.093& 0.118 & & 0.163 & 0.149 & 0.233\\
0.4 & 2 & 0.668 & 0.984 & 0.999 & & 0.806 &0.990 & 0.999\\
0.8 & 2 & 0.089 &0.168 & 0.324 & & 0.183 &0.350 & 0.523\\
1.6 & 2 & 0.058 &0.073 & 0.060 & & 0.116 & 0.099 & 0.122\\
2.4 & 2 & 0.081 &0.093 & 0.093 & & 0.165 & 0.127 & 0.189\\
0.4 & 3 & 0.545 & 0.921 & 0.992 & & 0.681 &0.954 & 0.993\\
0.8 & 3 & 0.075 & 0.110 & 0.203 & & 0.177 & 0.259 & 0.398\\
1.6 & 3 & 0.076 &0.064 & 0.065& & 0.102  &0.102 & 0.120\\
2.4 & 3 & 0.070 & 0.069 & 0.086 & & 0.146 & 0.121 & 0.185\\
\hline\hline
\end{tabular}
\end{center}
\caption{Simulated power for different configurations of $\Delta$, $\sigma^{2}$, $\alpha$, and $n_{\ell}$ in Scenario 4.}
\label{tab6MED}
\end{table}

\section{Conclusions}
\label{sec:conc}

In this paper, we used the results from \cite{Liu09} to derive a confidence interval for the maximum difference between two given non-linear regression models over the entire covariate space of interest and considered asymptotic methods to derive confidence intervals for the difference between two same target doses. One reviewer suggested comparing location and shape of the dose response curves  as a third way of assessing similarity in dose finding trials, in addition to the proposed assessments of similarity in dose response and target doses. One limitation of such an approach could be that even if two dose response functions have exactly the same location and shape, their curves could be very different. Consider, for example, the two Emax curves given by $m_1(d)=1+6d/(0.2+d)$ and $m_2(d)=1+6d/(3+d)$. Plotting those two curves, one recognizes immediately that they are very different although location and shape are the same. Building upon this comment, however, one possibility would be to construct a multivariate equivalence test on the entire model parameter vector. One could consider testing, for example, the null hypothesis $\tilde{H}\colon\|\vartheta_1-\vartheta_2\|_2^2 \geq \delta$ against the alternative hypothesis $\tilde{K}\colon\|\vartheta_1-\vartheta_2\|_2^2 < \delta$ for small values of $\delta$. Such a multivariate equivalence test could only be constructed if the two regression models are the same, which is a limitation compared to the methods proposed in this paper. In addition, non-parametric approaches could be considered such as the empirical probability plots presented by \cite{Doksum74,Doksum76}. Such approaches, however, may not readily apply to the situations considered in this paper and one would have to extend the theory considerably.

The choice of the equivalence margins $\delta$ and $\eta$ in \eqref{H0} and \eqref{hypothesesMED}, respectively, is a delicate problem.
This choice depends on the particular application and has to be made by clinical experts, possibly with input from statisticians and other quantitative scientists.
Regulatory guidance documents are available in specific settings, such as for the problem of demonstrating bioequivalence.
For example, \cite{EMA2010} discusses how the thresholds for bioequivalence hypotheses of the form considered
in this paper can be defined in various settings. For the comparison of curves as
considered in this paper we refer to Appendix 1 of \cite{EMA2010}, with emphasis on dissolution
profiles on the basis of specific measures.

A practical concern for any clinical trial is to determine its appropriate sample size. Related calculations at the trial design stage of a regular Phase II dose finding trial could be based either on power considerations to achieve a pre-specified probability of establishing a true dose response signal or on a pre-specified precision for dose response and target dose estimation. Using the methods proposed in this paper, sample size calculations will be based on testing the hypotheses $H, H'$, and $H''$ in the previous sections, such that the desired probability of rejecting the true null hypothesis under an assumed dose response curve is achieved. One may therefore justify the sample size using power calculations, with simulations performed to ensure adequate performance (as illustrated in Sections~\ref{sssec:pow} and~\ref{sssec:pow2}). In practice, the requirements for demonstrating similarity of target doses or dose response curves are more relaxed, not least because we may not have enough patients in the individual subgroups.

A particular application area of the proposed methods are multiregional clinical trials, which are run in different countries or regions, and potentially serve different submissions \cite{ICHE17}. For example,
many pharmaceutical companies focus on running multiregional
clinical trials that include a major Japanese subpopulation for later
regulatory submission in Japan. A natural question is then whether
the dose response results for the Japanese and the non-Japanese
populations are consistent \cite{Malinowski2008,Uesaka2009}.
For this purpose, the ICH E5 guideline \cite{ICHE5} recommends a separate trial to compare the dose response relationships between the two regions.
This document triggered considerably research (see \cite{Kawai00,Chow02,Goto02} among many others), although most literature at that time focused on bridging trials which is different to the situation considered here. Acknowledging that the comparison between two regions is not just a subgroup analysis problem \cite{ICHE17}, it would be interesting to see to which extent the proposed methods remain applicable in multiregional clinical trials.

In this paper, we focused on the comparison of two, possibly different, dose response models $m_1$ and $m_2$. In many situations limited information about the shape of the dose response curve is available at the trial design stage. For example, information might be available about the dose response curve for a similar compound in the same indication or the same compound in a different indication. Also, dose exposure response models might have been developed based on earlier data (e.g. from a proof-of-concept trial). Such information can be used for the clinical team to agree on a candidate dose response model. While empirical evidence suggests that the Emax model is often observed in clinical dose finding trials \cite{Thomas17,Li17}, model uncertainty remains of practical concern and is often underestimated. Especially, in the context of subgroup analysis and multiregional clinical trials the dose response models could be very different between subgroups and regions, respectively. Selecting a single model discards model uncertainty, which may lead to confidence intervals with a coverage probability smaller than the nominal level. Thus, alternative model selection and model averaging approaches have been investigated in the context of dose finding in Phase II \cite{Schorning16}, including the MCP-Mod approach \cite{Bretz05,EMA2014,FDA2016}. We leave the extension of the methods proposed in this paper to situations facing model uncertainty for future research.
\\


{\bf Acknowledgements}
This work has been supported in part by the Collaborative Research
Center "Statistical modeling of nonlinear dynamic processes" (SFB 823, Project C1) of the
German Research Foundation (DFG). Kathrin M\"ollenhoff's research has received funding
from the European Union Seventh Framework Programme [FP7 2007{2013] under grant
agreement no 602552 (IDEAL - Integrated Design and Analysis of small population group
trials). The authors would like to thank Georgina Bermann for many helpful discussions and bringing this case study to our attention.
They are also grateful to two referees and an associate editor for their valuable comments, which improved the presentation of this paper.

\section*{A. Coverage probability of the confidence interval for the maximum absolute difference }\label{appendixA}
\HD{
In the following we prove equation~\eqref{conf_band} from Section~\ref{ssec:meth1}.
To this end, let $d_0\in \cal D$ such that
$$\max_{d\in{\cal D}} |m_{2}(\mathbf{\vartheta}_{2},d)-m_{1}(\mathbf{\vartheta}_{1},d)|=|m_{2}(\mathbf{\vartheta}_{2},d_0)-m_{1}(\mathbf{\vartheta}_{1},d_0)|.$$
Hence
\begin{eqnarray}
\nonumber
{\cal P} &=&P\Big\{ \max_{d\in{\cal D}} |m_{2}(\mathbf{\vartheta}_{2},d)-m_{1}(\mathbf{\vartheta}_{1},d)|
\leq \max \big\{ \max_{d\in{\cal D}} U\left(Y_{1},Y_{2},d\right),-\min_{d\in{\cal D}} L\left(Y_{1},Y_{2},d\right)\big\} \Big\}\\
&=& \nonumber
 P\Big\{ |m_{2}(\mathbf{\vartheta}_{2},d_0)-m_{1}(\mathbf{\vartheta}_{1},d_0)|
\leq \max \big\{ \max_{d\in{\cal D}} U\left(Y_{1},Y_{2},d\right),-\min_{d\in{\cal D}} L\left(Y_{1},Y_{2},d\right)\big\} \Big\}\\
\nonumber &\geq&
P\Big\{ |m_{2}(\mathbf{\vartheta}_{2},d_0)-m_{1}(\mathbf{\vartheta}_{1},d_0)|
\leq \max \big\{ U\left(Y_{1},Y_{2},d_0\right),-L\left(Y_{1},Y_{2},d_0\right)\big\} \Big\}.
\end{eqnarray}
Now we distinguish two cases. If $m_{2}(\mathbf{\vartheta}_{2},d_0)-m_{1}(\mathbf{\vartheta}_{1},d_0)\geq 0$ we have
\begin{eqnarray} \label{a1}
{\cal P}  &\geq&
 P\Big\{ m_{2}(\mathbf{\vartheta}_{2},d_0)-m_{1}(\mathbf{\vartheta}_{1},d_0)
\leq U\left(Y_{1},Y_{2},d_0\right)\Big\} \stackrel{n_1,n_2\rightarrow \infty} \longrightarrow  1-\alpha,
\end{eqnarray}
as $U\left(Y_{1},Y_{2},d\right)$ is a $1-\alpha$ pointwise upper confidence bound on $m_{2}(\mathbf{\vartheta}_{2},d)-m_{1}(\mathbf{\vartheta}_{1},d)$.
Otherwise, $m_{2}(\mathbf{\vartheta}_{2},d_0)-m_{1}(\mathbf{\vartheta}_{1},d_0)\leq 0$ and the same argument applies to $L\left(Y_{1},Y_{2},d\right)$, yielding
\begin{eqnarray}
{\cal P} & {\geq}&
 P\Big\{ m_{2}(\mathbf{\vartheta}_{2},d_0)-m_{1}(\mathbf{\vartheta}_{1},d_0) \geq L\left(Y_{1},Y_{2},d_0\right)\Big\}
\stackrel{n_1,n_2\rightarrow \infty}  \longrightarrow  1-\alpha.
 \label{a2}
\end{eqnarray}
}
\section*{B. Asymptotic level of  the test for similarity of two target doses}\label{appendixB}

\HD{
We show that the test \eqref{testMED} defined in Section~\ref{ssec:meth2} has asymptotic level $\alpha$, that is
\begin{equation}
    \lim_{n_1,n_2\rightarrow \infty} P\left(|\widehat{MED}_1-\widehat{MED_2}|\leq c\right)\leq\alpha
    \label{levelMED}
\end{equation}
under the null hypothesis.
First note  that the solution of  equation \eqref{constant_c}  is unique as the function $c \rightarrow \Phi\left(\frac{c-\eta}{\hat{\tau}}\right)-\Phi\left(\frac{-c-\eta}{\hat{\tau}}\right)$
is strictly increasing  with limits $-1$ and $1$ as $c \rightarrow  - \infty $ and $\infty$, respectively.
Next, let $t=MED_1-MED_2$, $\hat t =\widehat{MED}_1-\widehat{MED_2} $ and denote the power function of the test by
\begin{equation*}
    G_{n_1,n_2}(\theta)=P\left(\left|\hat{t}\right|<c\right).
\end{equation*}
 The assertion \eqref{levelMED} is then equivalent to
\begin{equation}
    \lim_{n_1,n_2\rightarrow \infty} G_{n_1,n_2} (\theta)\leq\alpha\ \quad \mbox{for all } |t| \geq \eta
    \label{power}.
\end{equation}
A standard calculation shows that
\begin{eqnarray*}
    G_{n_1,n_2}(t)&=&P (|\hat{t}|\leq c)=P(-c\leq\hat{t}\leq c)=P \Big (\frac{-c-t}{\hat{\tau}}\leq\frac{\hat{t}-t}{\hat{\tau}}\leq \frac{c-t}{\hat{\tau}}     \Big ) \\
&& \stackrel{n_1,n_2\rightarrow \infty}\longrightarrow   \tilde{G}(t)  :=  \Phi  \Big (\frac{c-t}{\tau} \Big  )-\Phi  \Big  (\frac{-c-t}{\tau}  \Big )
\end{eqnarray*}
Now consider the problem of testing the hypotheses $H\colon\left|t\right|
\geq\eta$ against   $K\colon\left|t\right|<\eta$ for normally distributed data $X  \sim\mathcal{N}(t, \tau^2)$.  A simple calculation shows
that  the (asymptotic) power function $\tilde{G}$ coincides with the power of the test,
which rejects the null hypothesis $H\colon\left|t\right| \geq\eta$ whenever  $|X|  \leq c$.
Considering  the discussion in Lehmann et al. \cite[p. 81]{lehmann}, it follows that this test is  uniformly most powerful and unbiased of
 size $\alpha$.
This implies  $\tilde{G}(t) \leq \tilde{G}(\eta)=\alpha$ for all $|t| \geq \eta$ and proves  \eqref{power}.
}

\end{document}